\documentclass[a4paper,11pt]{article}
\usepackage{jheppub_modified_green} 
\usepackage{tensor}
\usepackage{mathtools}
\usepackage{subcaption}
\usepackage{booktabs}
\usepackage{bbold}
\usepackage{pgfplots}
\pgfplotsset{compat=1.14}
\usepackage{cancel}
\usepackage{soul}
\usepackage{mathabx}
\usepackage{xcolor}
\usepackage{tikz}
 \usepackage[normalem]{ulem} 
\usepackage{tikz-3dplot}
\usetikzlibrary{hobby,calc,shapes, positioning, backgrounds,trees, decorations, decorations.markings, decorations.pathmorphing, arrows, shapes.geometric, snakes,arrows.meta, angles, quotes,tikzmark}
\tikzset{
	graviton/.style={line width=.8pt, -latex,decorate, decoration={snake, segment length=4pt,amplitude=1.8pt, pre length=.1cm, post length=.25cm}},
	worldline/.style={gray, line width=1pt},
	worldlineBold/.style={black, line width=.6pt},
	zUndirected/.style={line width=1pt},
	zParticle/.style={line width=1pt,postaction={decorate},decoration={markings,mark=at position .6 with {\arrow[#1]{latex}}}},
	zParticleF/.style={line width=1pt,postaction={decorate}},
	cscalar/.style={line width=1pt,postaction={decorate},decoration={markings,mark=at position .6 with {\arrow[#1]{latex}}}},
	cscalar2/.style={line width=1pt,postaction={decorate},decoration={markings,mark=at position .8 with {\arrow[#1]{latex}}}},
	photon/.style={line width=.8pt, decorate, decoration={snake, segment length=4pt, amplitude=1.8pt,  pre length=.1cm, post length=.1cm}},
    photon2/.style={line width=2pt, gray, decorate, decoration={snake, segment length=5pt, amplitude=1.8pt,  pre length=.1cm, post length=.1cm,}},
    cross/.style={path picture={ 
      \draw[gray,thick]
    (path picture bounding box.south east) -- (path picture bounding box.north west) (path picture bounding box.south west) -- (path picture bounding box.north east);
    }}
}

\usepackage{tikz-feynman}
\tikzfeynmanset{compat=1.1.0}
\tikzfeynmanset{
graviton/.style={circle, draw=green!60, fill=green!5, very thick, minimum size=7mm}
}
\tikzfeynmanset{
codot/.style={/tikz/shape=circle,
/tikz/fill=red,/tikz/minimum size=0.1cm,/tikz/inner sep=1.8pt}
}
\tikzfeynmanset{sb/.style=
{/tikz/shape=circle,
/tikz/fill=black,
 /tikz/minimum size=0.3cm,/tikz/inner sep=1.pt
 } }
\tikzfeynmanset{myblob/.style=
{/tikz/shape=ellipse,
/tikz/fill=red,
 /tikz/minimum width=0.5cm,
 } }
 \tikzfeynmanset{myblobFF/.style=
{/tikz/shape=circle,
/tikz/fill=black,
 /tikz/minimum width=0.25cm,
 } }
 \tikzfeynmanset{myblob2/.style=
{/tikz/shape=rectangle,
/tikz/fill={rgb:black,1;white,2},
 /tikz/minimum width=0.01cm,/tikz/minimum height=0.3cm,/tikz/inner sep=0.1pt
 } }
  \tikzset{GR/.style={
  ellipse,
  fill={rgb:black,1;white,2},
  minimum size=0.1cm,
  draw=none
}}
  \tikzfeynmanset{fp/.style=
{/tikz/shape=rectangle,
/tikz/fill=red,
 /tikz/minimum width=0.1cm,/tikz/inner sep=1.8pt
 } }
  \tikzfeynmanset{myvertex/.style=
{/tikz/shape=circle,
/tikz/fill=black,
 /tikz/minimum width=0.05cm,
 } }
  \tikzfeynmanset{myvertex1/.style=
{/tikz/shape=circle,
/tikz/fill=gray,
 /tikz/minimum width=0.05mm,
 } }
\tikzset{box/.pic={\filldraw[fill=black]  (0,0) circle (2.5pt);
				   \filldraw [fill=black] (0.5,0) circle (2.5pt);
			       \draw [line width=5pt] (0,0) -- (0.5,0);}}

\tikzset{wiggle/.style={decorate, decoration=snake}}

 \tikzfeynmanset{HV/.style=
{/tikz/shape=circle,
/tikz/fill={rgb:black,1;white,2},
 /tikz/minimum size=0.01cm,/tikz/inner sep=1.8pt
 } }
\tikzfeynmanset{HV2/.style=
{/tikz/shape=circle,
/tikz/fill={rgb:black,1;white,2},
 /tikz/minimum size=0.01cm,/tikz/inner sep=0.1pt
 } }

\newcommand{\black}{\color{black}}

\def\euler{\mathrm{e}}

\newcommand{\npre}{\mathcal{N}}

\newcommand{\mdot}{{\cdot}}
\newcommand{\eps}{\epsilon}

\def\veps{\varepsilon}

\def\nn{\nonumber}
\usepackage{hyperref}
\hypersetup{
	colorlinks=true,
	linktoc=page,
	citecolor=americanrose,
	linkcolor=cadmiumgreen,
	urlcolor=blue} 
\definecolor{americanrose}{rgb}{1.0, 0.01, 0.24}
\definecolor{cadmiumgreen}{rgb}{0.0, 0.42, 0.24}

\newcommand{\CN}{\mathcal{N}}
\newcommand{\CM}{\mathcal{M}}

\newcommand{\rmS}{>}
\newcommand{\rmA}{<}
\newcommand{\XX}{\mathcal{X}}
\newcommand{\bS}{\mathbb{S}}
\newcommand{\bG}{\mathbb{\Gamma}}
\newcommand{\bT}{\mathbb{T}}

\newcommand{\terms}[1]{\mathcal{O}\left(#1\right)}
\newcommand{\epsbar}{\bar\eps}

\title{ 
The Effective Kerr-Compton Tree Amplitude from Black Hole Perturbation Theory
}

\author{N. Emil J. Bjerrum-Bohr$^{1,2}$,}
\author{Gang Chen$^{1,2}$,}
\author{Konstantinos Papadimos$^{3,4}$,}
\author{Nabha Shah$^{1,2}$,}
\author{Chenliang Su$^{2}$}
\affiliation{$^1$Center of Gravity, Niels Bohr Institute, University of Copenhagen,
Blegdamsvej 17, DK-2100 Copenhagen, Denmark}
\affiliation{$^2$Niels Bohr International Academy, Niels Bohr Institute, University of Copenhagen,\\
Blegdamsvej 17, DK-2100 Copenhagen, Denmark}
\affiliation{$^3$Max Planck Institute
for Gravitational Physics, Albert Einstein Institute,\\
Am M\"uhlenberg 1,
14476 Potsdam, Germany}
\affiliation{$^4$Institut f\"ur Physik und IRIS Adlershof, Humboldt-Universit\"at zu Berlin, \\
Zum Gro\ss en Windkanal 2, 
D-12489 Berlin, Germany}

\emailAdd{bjbohr@nbi.dk}
\emailAdd{gang.chen@nbi.ku.dk}
\emailAdd{konstantinos.papadimos@aei.mpg.de}
\emailAdd{nabha.shah@nbi.ku.dk}
\emailAdd{chenliang.su@nbi.ku.dk}

\abstract{
We develop a new method to construct the tree-level Compton amplitude for gravitational-wave scattering off a Kerr black hole in post-Minkowskian effective field theory. Building on previous computations that required physical consistency of the amplitude, the method uses wave scattering functions from black hole perturbation theory to uniquely determine additional contact terms. We illustrate the iterative procedure by deriving explicit expressions up to sixth order in spin, including previously unknown sectors at the fifth and sixth orders in spin. The amplitude features complex transcendental structures and a leading dissipative contact term at sixth order in spin. We verify our result against known results in the literature and reproduce the one-loop tail result from black hole perturbation theory.}


\begin{document}

\maketitle

\tableofcontents

\flushbottom

\newpage

\section{Introduction}
Particle physics approaches to the analysis of classical gravitational systems, as pioneered in~\cite{Iwasaki:1971iy}, typically start from a point-like approximation that essentially treats a massive body or a black hole as a fundamental particle. Naturally suited to the study of scattering massive states, such effective descriptions can be valuable for understanding two-body interactions and informing the dynamics of coalescing binary black holes in the inspiral phase, for instance, in the context of the effective-one-body formalism~\cite{Buonanno:1998gg, Damour:2017zjx}. When modeling the scattering of non-spinning black holes in a weak-field or post-Minkowskian expansion, the point-like approximation is exact at low perturbative orders. However, it breaks down with the emergence of tidal effects at higher orders or through the inclusion of spin and needs systematic correction. Notable progress has been made in understanding compact object scattering involving finite-size effects by employing various field-theoretic methods. Worldline effective field theory techniques~\cite{Goldberger:2004jt,Goldberger:2005cd, Porto:2005ac,Porto:2007qi} provide a systematic framework for representing internal structure. In parallel, modern quantum field theory tools offer powerful methods for encoding finite-size interactions in on-shell scattering amplitudes from which classical gravitational dynamics and observables can be extracted. These approaches build on the developments in~\cite{Neill:2013wsa, Bjerrum-Bohr:2013bxa,Bjerrum-Bohr:2018xdl,Cheung:2018wkq,Kosower:2018adc,Brandhuber:2021eyq} and have been further advanced in, {\it e.g.,} Refs.~\cite{Chung:2020rrz,Akpinar:2024meg,Bautista:2024emt,Vines:2017hyw,Arkani-Hamed:2017jhn,Guevara:2018wpp,Chung:2018kqs,Guevara:2017csg,Vines:2018gqi,Guevara:2019fsj,Arkani-Hamed:2019ymq,Guevara:2020xjx,Kosmopoulos:2021zoq,Bern:2020buy,Jakobsen:2021zvh,Maybee:2019jus,Haddad:2021znf,Liu:2021zxr,Jakobsen:2022fcj,Menezes:2022tcs,FebresCordero:2022jts,Alessio:2022kwv,Aoude:2022thd,Damgaard:2022jem,Aoude:2023vdk,Bianchi:2023lrg,Bautista:2023szu,Akpinar:2025bkt,Bjerrum-Bohr:2026fhx,Brandhuber:2023hhl,DeAngelis:2023lvf,Aoude:2023dui,Bohnenblust:2023qmy,Aoude:2022trd,Chen:2022kpm,Bern:2022kto}. 

An interesting theoretical component in such computations is the classical gravitational Compton amplitude~\cite{Cangemi:2023bpe,Scheopner:2023rzp,Bautista:2024agp,Aoude:2020onz,Chiodaroli:2021eug,Bautista:2021wfy,Bautista:2022wjf,Bjerrum-Bohr:2023jau,Bjerrum-Bohr:2023iey,Chen:2022clh,Chen:2024mlx,Correia:2024jgr,Vazquez-Holm:2025ztz,Ivanov:2022qqt,Cangemi:2022bew,Saketh:2024juq,Akpinar:2025huz,Akpinar:2025byi,Caron-Huot:2025tlq,Bjerrum-Bohr:2025lpw,Correia:2026utp,Ochirov:2022nqz,Bjerrum-Bohr:2026glv,Bautista:2026qse,Ivanov:2026icp,Brunello:2026rdk,Bautista:2026fcp,Brunello:2026lzf}, which describes the scattering of a graviton off a massive source and captures information on the source's leading linear finite-size effects. As a building block in loop-level amplitudes for the interaction of two extended massive bodies, the Compton amplitude enables systematic perturbative calculation of observables.

The computation of the Kerr-Compton amplitude in a post-Minkowskian effective field theory begins with the three-point amplitude involving a graviton and a massive particle with classical spin. Using input from studies of the Kerr solution~\cite{Kerr:1963ud}, one can fix the on-shell three-point amplitude to arbitrary spin order~\cite{Vines:2017hyw} and verify that its resummation takes an exponential form~\cite{Arkani-Hamed:2017jhn}, making it particularly suitable for capturing nonperturbative spin effects~\cite{Guevara:2018wpp,Chung:2018kqs,Chen:2022kpm}. An exploration of how to use these results to derive the Compton amplitude appeared in~\cite{Bautista:2021wfy,Bautista:2022wjf}, where the tree-level amplitude was derived from the far-zone component of the exponential phase shift of the gravitational wave in the Mano-Suzuki-Takasugi solution~\cite{Mano:1996vt}. This methodology and other approaches outlined in references~\cite{Chen:2022kpm,Bern:2022kto,Aoude:2022trd,Bjerrum-Bohr:2023iey,Cangemi:2023bpe,Scheopner:2023rzp,Bautista:2024agp} agree on results to quartic order in spin. 

In this paper, we derive the tree-level effective Kerr-Compton amplitude at higher orders in spin using a general and systematic procedure for extracting finite-size contributions directly from the wave scattering functions obtained from black hole perturbation theory. This method bypasses the need for near- and far-zone separation, and allows us to work directly in the physical black hole spin region $|a|<GM$, where $M$ is the black hole mass, $G$ is Newton's constant, and $|a|$ denotes the spin length. To extract the effective amplitude, a projection onto partial waves is required, but it is ill-defined due to the leading singularities in the forward scattering limit. Following~\cite{Bjerrum-Bohr:2026glv}, we resolve this issue by rewriting the amplitude as a resummed Newtonian contribution, which regularizes the divergence, and a remainder free of such singularities. This regularization allows us to fix global phase conventions and establish the equivalence between the partial-wave components of the amplitude in the effective field theory and in black hole perturbation theory. In our procedure, we account for the fact that tree-level finite-size contributions cannot be extracted directly at a given post-Minkowskian order, since horizon-flux contributions associated with a real black hole’s spin~\cite{Hawking:1972hy,Starobinskii:1973vzb,Teukolsky:1974yv} involve negative powers of $GM$, whereas tidal deformations involve positive powers of $GM$.

The paper is organized as follows. In Section~\ref{sec:ampstructure}, we review the structure of the classical Kerr-Compton amplitude in effective field theory and introduce the corresponding  results in  black hole perturbation theory. We then outline the procedure to construct  contact  terms in the tree-level effective Compton amplitude in Section~\ref{sec:TreeKerr}. Section~\ref{sec:contactTerms} describes the application of the outlined procedure to extract the complete tree-level effective amplitude at fifth and sixth orders in spin, after which we conclude with final remarks and directions for further research.

\section{Structure of the Kerr-Compton Amplitude}
\label{sec:ampstructure}
We begin by revisiting the tree-level amplitude in post-Minkowskian effective field theory for interactions between a Kerr black hole and a massless graviton, following the analysis in~\cite{Bjerrum-Bohr:2023iey,Bjerrum-Bohr:2023jau}. The process of interest is represented as
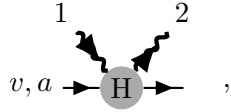
\begin{equation*}\label{fig:Compton amplitude}
\centering
\begin{tikzpicture}[scale=1.,transform shape]
		\begin{feynman}
			\vertex (a) {\(v,a\)};
			\vertex [right=1.2cm of a] (f2) [HV] {H};
			\vertex [right=1.2cm of f2] (c) {\, \, ,};
			\vertex [above=1.cm of f2] (gm) {};
			\vertex [left=0.8cm of gm] (g2) {$1$};
			\vertex [right=0.8cm of gm] (g20) {$2$};
			
			\diagram* {
				(a) -- [fermion,thick] (f2) -- [fermion,thick] (c),
				(g2) -- [
				photon,
				ultra thick,
				postaction={
					decorate,
					decoration={
						markings,
						mark=at position 0.8 with
						{\arrow[scale=1.3, very thick]{Latex}}
					}
				}
				] (f2),
				(f2) -- [
				photon,
				ultra thick,
				postaction={
					decorate,
					decoration={
						markings,
						mark=at position 0.7 with
						{\arrow[scale=1.3, very thick]{Latex}}
					}
				}
				] (g20)
			};
		\end{feynman}
	\end{tikzpicture}
\end{equation*}
where the incoming and outgoing graviton legs carry on-shell momenta $p_1^\mu$ and $p_2^\mu$ respectively. Without loss of generality, their polarization tensors are expressed as products of vector polarizations $\veps_1^\mu\veps_1^\nu$ and $\veps_2^\mu\veps_2^\nu$. The Kerr black hole has mass $M$, velocity $v^\mu$, and spin $a^\mu$, with $a\mdot v=0$. The commonly used spin tensor is related to the spin vector via $S^{\mu\nu}=M\epsilon^{\mu\nu\alpha\beta}v_\alpha a_\beta$, where $\epsilon^{\mu\nu\alpha\beta}$ is the Levi-Civita tensor, under the spin supplementary condition $S^{\mu\nu}v_\nu=0$~\cite{tulczyjew1959motion,Dixon:1970zza}. 

In the classical limit, the incoming and outgoing gravitons carry the same energy $ \omega=p_1\mdot v=p_2\mdot v$, and the relevant contribution to the amplitude arises solely from the $\mathcal{O}(M^2)$ term in the heavy-mass expansion; see~\cite{Brandhuber:2021eyq} for details. The main result of~\cite{Bjerrum-Bohr:2023jau} is that the tree-level spinning Compton amplitude can be expressed as
\begin{equation}\label{eq:allSpinCompTree}
       \CM^{\text{tree}}_{\text{Kerr}}=32\pi G M^2\Big( \frac{\CN_0\CN_a}{2p_1\mdot p_2 }-\frac{\CN_{\rm r}}{4\, p_1\mdot v \, p_2\mdot v}+\CN_{\rm c}\Big).
\end{equation}
The first term is the double-copy contribution~\cite{Bern:2008qj,Bern:2010ue,Bern:2019prr} obtained by multiplying a gauge-theory numerator $\CN_a$ with classical spin $|a|$ and its spinless counterpart $\CN_0$. This term reproduces the correct amplitude factorization associated with the massless pole. The numerator $\CN_0$, which can be derived from the kinematic Hopf algebra \cite{Brandhuber:2021kpo,Brandhuber:2021bsf,Brandhuber:2022enp,Chen:2022nei,Chen:2023ekh,Chen:2024gkj,Bjerrum-Bohr:2024fbt}, is given by
\begin{align}
	\npre_0=\!\frac{v\mdot F_1\mdot F_2^*\mdot v}{p_1\mdot v}\,,
\end{align}
where
$F_i^{\mu\nu}=p_i^{\mu}\veps_i^{\nu}-\veps_i^{\mu}p_i^{\nu}$ denotes the abelian field-strength tensor and the asterisk
denotes complex conjugation. The gauge-theory numerator $\CN_a$ can be expressed in terms of
\begin{equation}
    G_1(a\mdot p_i) \equiv \frac{\sinh{a\mdot p_i}}{a\mdot p_i}\,,
\end{equation} 
and 
\begin{equation}
	G_2(a\mdot p_1, a\mdot p_2)\equiv {1\over a\mdot p_{2}}\Big({\sinh(a\mdot p_1+a\mdot p_2)\over a\mdot p_1+a\mdot p_2}-\cosh(a\mdot p_2) {\sinh(a\mdot p_1)\over a\mdot p_1}\Big)\,,
\end{equation} 
so that
\begin{align}\label{eq:NaDC}
	&\npre_a=\frac{{\mathsf w}(p_1)\mdot F_1\mdot F_2^*\mdot {\mathsf w}(-p_2)}{M^2 (p_2\mdot v)}\\
    &+\frac{i}{M} \Big( G_2\left(a\mdot p_1,-a\mdot p_2\right) (a\mdot F_1\mdot F_2^*\mdot S\mdot p_2-a\mdot F_2^*\mdot F_1\mdot S\mdot p_1)- G_1\left(a\mdot p_1-a\mdot p_2\right) \text{tr}\left(F_1\mdot S\mdot F_2^*\right)\Big) \nn\\
      &- G_1(a\mdot p_1) G_1(-a\mdot p_2) \bigg( (a\mdot F_1\mdot v) (a\mdot F_2^*\mdot p_1)\!+\!(a\mdot F_1\mdot p_2) (a\mdot F_2^*\mdot v) +\!\frac{p_1\mdot v\!+\!p_2\mdot v}{2}  (a\mdot F_1\mdot F_2^*\mdot a)\bigg)\,,\nn
\end{align}
with ${\mathsf w}^{\mu}(p_i)=M v^{\mu}\cosh(a\mdot p_i)-i\,\,(p_i\mdot S)^{\mu}\,G_1(a\mdot p_i)$. Because $G_1$ and $G_2$ are entire functions, $\npre_a$ is manifestly free of unphysical poles at
$a\mdot p_1=0$, $a\mdot p_2=0$, and $a\mdot p_1-a\mdot p_2=0$.

The second term in eq.~\eqref{eq:allSpinCompTree}, referred to as the remainder contribution, is fixed by the factorization behavior on the massive propagator pole~\cite{Chen:2022kpm}. Defining $\widetilde F^{\mu\nu}={1\over 2}\epsilon^{\mu\nu\rho\sigma}F_{\rho\sigma}$, we can write the numerator as
\begin{align}
	 \npre_{\rm r}=& (\partial_{a\mdot p_1}+\partial_{a\mdot p_2})\Big(G_1(a\mdot p_1)G_1(-a\mdot p_2)\Big) \nn\\
     &\times \bigg( p_2\mdot v \big((a\mdot F_1\mdot F_2^*\mdot a) (a\mdot F_2^*\mdot F_1\mdot v)+ a\mdot a (a\mdot F_1\mdot F_2^*\mdot v) (v\mdot F_1\mdot F_2^*\mdot v)\big)+(1\leftrightarrow2) \bigg)\nn\\
     &+i(\partial_{a\mdot p_1}+\partial_{a\mdot p_2})G_2(a\mdot p_1,-a\mdot p_2)\nn\\
     &\times\bigg( p_1\mdot v(a\mdot F_1\mdot F_2^*\mdot v)\big((a\mdot F_1\mdot v)(a\mdot \widetilde F_2^*\mdot v)-(a\mdot F_2^*\mdot v)(a\mdot \widetilde F_1\mdot v)\big) -(1\leftrightarrow 2)\bigg)\, .
\end{align} 
Finally, the amplitude contains contact terms $\CN_{\rm c}$ that cannot be fully determined by physical factorization consistency. These terms encode additional information about the interaction between a graviton and a Kerr black hole. In this paper, we outline a procedure to uniquely fix $\CN_{\rm c}$ by matching the amplitude to the corresponding results in black hole perturbation theory.

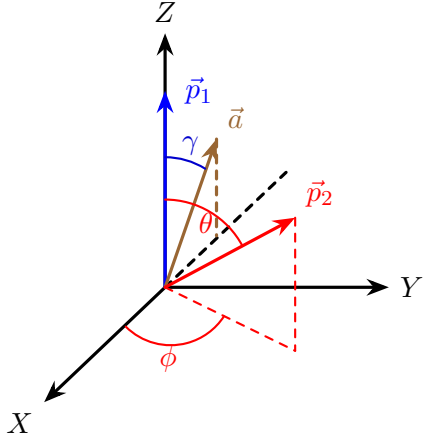
\begin{figure}[h]
\centering
\begin{tikzpicture}[
    >=Stealth,
    line cap=round,
    line join=round,
    scale=0.8
]

\coordinate (O) at (0,0);

\draw[->, very thick] (O) -- (0,4.2)
    node[above] {$Z$};

\draw[->, very thick] (O) -- (3.7,0)
    node[right] {$Y$};

\draw[->, very thick] (O) -- (-2.0,-1.9)
    node[below left] {$X$};
\draw[dashed, very thick] (O) -- (2.0,1.9)
    ;
\draw[->, very thick, blue]
    (O) -- (0,3.25)
    node[right=0.12cm] {$\vec p_1$};

\coordinate (A) at (0.85,2.45);

\draw[->, very thick, brown!80!black]
    (O) -- (A)
    node[above right] {$\vec a$};
\draw[dashed, very thick, brown!80!black]
    (A) -- (0.85,0.85);

\coordinate (P) at (2.15,1.15);

\draw[->, very thick, red]
    (O) -- (P)
    node[above right] {$\vec p_2$};

\coordinate (Q) at (2.15,-1.05);

\draw[dashed, red, thick]
    (P) -- (Q);

\draw[dashed, red, thick]
    (O) -- (Q);

\draw[
    blue!80!black,
    thick,
    -,
    domain=0:0
] (0,2.15) arc[start angle=90,end angle=58,radius=1.25];

\node[blue!80!black] at (0.42,2.35) {$\gamma$};

\draw[
    red,
    thick,
    -,
    domain=0:0
] (0,1.45) arc[start angle=90,end angle=28,radius=1.45];

\node[red] at (0.68,1.07) {$\theta$};

\draw[
    red,
    thick,
    -,
    domain=0:0
] (-0.65,-0.65)
    arc[start angle=225,end angle=327,radius=1.05];

\node[red] at (0.05,-1.18) {$\phi$};

\end{tikzpicture}
\caption{Kinematic configuration for the Compton scattering process under consideration. With this specification, the massless graviton propagator has the simple dependence $\sim1/\sin^2(\theta/2)$, which is particularly convenient for projection onto the spherical-harmonic basis.}
\label{fig:XYZ}
\end{figure}

To perform the comparison, we work in the black hole's rest frame and adopt the coordinates and configuration shown in Figure~\ref{fig:XYZ}. The incoming spatial momentum $\vec p_1$ is aligned with the $Z$-axis, and the three-dimensional spin vector $\vec a$ lies in the $X\!-\!Z$ plane, with $\gamma$ denoting its inclination relative to the $Z$-axis. The outgoing spatial momentum $\vec p_2$ is parametrized by the polar and azimuthal angles $(\theta,\phi)$. With this convention, we specify 
\begin{align}\label{eq:cood}
	 v&=(1,0,0,0)\,,\quad p_1=\omega(1,0,0,1 )\,,\quad
   \varepsilon^{+}_1=\frac{1}{\sqrt{2}}(0,1,i  ,0)\,,\quad
   \varepsilon^{-}_1=\frac{1}{\sqrt{2}}(0,1  ,
  -i ,0 )\,,  \nn\\
  a \!  &=|a|(0,-\sin(\gamma),0,\cos(\gamma))\,, 
  \qquad p_2=\omega  (1,\sin (\theta) \cos (\phi),\sin (\theta) \sin (\phi),\cos (\theta)),\,\nn\\
   \varepsilon^{+}_2&=\frac{1}{\sqrt{2}}(0,\cos (\theta ) \cos (\phi )-i \sin (\phi ),\cos (\theta ) \sin (\phi)+i \cos (\phi ),-\sin (\theta ))\,,\nn\\
   \varepsilon^{-}_2&=\frac{1}{\sqrt{2}}(0,\cos (\theta ) \cos (\phi )+i \sin (\phi ),\cos (\theta ) \sin (\phi)-i \cos (\phi ),-\sin (\theta )) \, ,
\end{align}
so that the scattering amplitude may be written as a function of the angles.

The relevant physical observable for the scattering process is the differential cross-section, which can be obtained from the finite part~\cite{Bjerrum-Bohr:2025bqg,JordanEriksen:2026lox,Bjerrum-Bohr:2026glv} of the effective-field-theory amplitude $\mathcal{M}_{\text{fin.}}(\mu=2e^{-\gamma_{\rm E}}\omega, \varepsilon_1, \varepsilon_2^{*})$, after removing the Weinberg phase~\cite{Weinberg:1965nx}. Following~\cite{Bjerrum-Bohr:2026glv}, we set the arbitrary mass scale arising from dimensional regularization of loop integrals to $\mu=2e^{-\gamma_{\rm E}}\omega$. The differential cross section is therefore given by
\begin{align}
{d\sigma\over d\Omega} = {1\over (8 \pi M)^2} \sum_{h=\pm}|\mathcal{M}_{\text{fin.}}(2e^{-\gamma_{\rm E}}\omega, \varepsilon^+_1, \varepsilon_2^{h*})|^2= |f_{\rm amp.}(\theta,\phi)|^2+ |g_{\rm amp.}(\theta,\phi)|^2 ,
\end{align}
with 
\begin{align}
 f_{\rm amp.}(\theta,\phi)&={1\over (8 \pi M)} \mathcal{M}_{\text{fin.}}(2e^{-\gamma_{\rm E}}\omega, \varepsilon^{+}_1, \varepsilon^{+*}_2),\nn\\
 g_{\rm amp.}(\theta,\phi)&={1\over (8 \pi M)} \mathcal{M}_{\text{fin.}}(2e^{-\gamma_{\rm E}}\omega, \varepsilon^{+}_1, \varepsilon^{-*}_2).
\end{align}
In black hole perturbation theory, the cross-section has an equivalent decomposition,
\begin{align}
	\frac{d\sigma}{d\Omega}
	=
	|f_{\rm B}|^2+|g_{\rm B}|^2,
\end{align}
into the helicity-conserving scattering function $f_{\rm B}$ and the helicity-reversing scattering function $g_{\rm B}$. These functions are obtained by solving the wave equation for linear perturbations of a Kerr metric and admit partial-wave expansions~\cite{Matzner:1977dn,Dolan:2008kf,Dolan:2007ut} in terms of spin-weighted spheroidal harmonics~\cite{Fackerell:1977},
\begin{align}
\label{eq:fBL}
f_B(\theta,\phi)
&=
\frac{\pi e^{-i\Phi}}{i\omega}\sum_{l=2}^{\infty}
\sum_{m=-l}^{l}
{}_{-2}\widehat{S}_{lm}(\theta,\phi;|a|\omega)\,
{}_{-2}S_{lm}(\gamma,0;|a|\omega)\Delta^{+}_{lm},
\\[1em]
\label{eq:gBL}
g_B(\theta,\phi)
&=\frac{\pi e^{-i\Phi}}{i\omega}
\sum_{l=2}^{\infty}
\sum_{m=-l}^{l}
{}_{-2}\widecheck{S}_{lm}(\pi-\theta,\phi;|a|\omega)\, {}_{-2}S_{lm}(\gamma,0;|a|\omega)\, (-1)^{l+m} \Delta^{-}_{lm},
\end{align}
where
\begin{align}
\Delta^{+}_{lm}&\equiv
\Big(
e^{2i\delta^{+}_{lm}}
+
e^{2i\delta^{-}_{lm}}
\Big),&
\Delta^{-}_{lm}&\equiv
\Big(
e^{2i\delta^{+}_{lm}}
-
e^{2i\delta^{-}_{lm}}
\Big)\, .
\end{align}
Here, ${}_{-2}S_{lm}$ are the spheroidal harmonic functions in the frame where the spin is aligned with the $Z$-axis and ${}_{-2}\widehat{S}_{lm}$ and $ {}_{-2}\widecheck{S}_{lm}$ denote the same objects expressed in the coordinate system shown in Figure~\ref{fig:XYZ}; see Appendix~\ref{app:BHPT} for details and explicit expressions. For a consistent comparison to the effective field theory amplitude, we follow~\cite{Bjerrum-Bohr:2026glv} and omit the subtraction of $2$ from the traditional expression for $\Delta^{+}_{lm}$. Furthermore, we introduce the overall phase $\Phi = 2 \bar \epsilon \ln(2\bar \epsilon)-\bar \epsilon$, which does not affect the physical cross section. The quantities $\delta_{lm}^{\pm}$ are the partial-wave phase shifts associated with the two parity sectors. They are determined from the Mano--Suzuki--Takasugi solution~\cite{Mano:1996vt} of the Teukolsky equation~\cite{Teukolsky:1973ha,Teukolsky:1972my} through
\begin{align}
e^{2i\delta_{lm}^{\pm}}
&=
(-1)^{l+1}
\frac{C_{\rm TS}}{16\omega^4}
\frac{B_{lm}^{\rm (ref)}}{B_{lm}^{\rm (inc)}},
\end{align}
where $C_{\rm TS}$ is the Teukolsky--Starobinsky constant, and $B_{lm}^{\rm (inc)}$ and $B_{lm}^{\rm (ref)}$ denote the incident and reflected wave coefficients, respectively. Their explicit values can be readily obtained using the efficient computational toolkit described in~\cite{BHPToolkit}. 

Agreement between the amplitudes in the two descriptions requires
\begin{align}\label{eq:compAll}
f_{\rm B}(\theta,\phi)
&=
f_{\rm amp.}(\theta,\phi),&
g_{\rm B}(\theta,\phi)
&=
g_{\rm amp.}(\theta,\phi). 
\end{align}
These equations provide complete constraints on $\CN_c$. In practice, it is more convenient to perform the comparison mode by mode in the partial-wave decomposition where the components of the amplitudes, or scattering functions, are defined via the projection integrals,
\begin{equation}\label{eq:Pfg}
\begin{aligned}
\mathbb{P}^{(l,m)}_f(\XX)&\equiv
\int_0^{\pi} \sin(\theta)d\theta \int_{0}^{2\pi} d\phi\,
{}_{-2}Y_{lm}^{*}(\theta,\phi)\,
\XX(\theta,\phi)\,,
\\
\mathbb{P}^{(l,m)}_g(\XX)&\equiv
\int_0^{\pi} \sin(\theta)d\theta \int_{0}^{2\pi} d\phi\,
{}_{-2}Y_{lm}^{*}(\pi-\theta,\phi)\,
\XX(\theta,\phi) \,,
\end{aligned}
\end{equation}
such that
\begin{align}
f_{\rm B}^{(l,m)}
&=\mathbb{P}^{(l,m)}_f(f_{\rm B}), &
g_{\rm B}^{(l,m)}
&=\mathbb{P}^{(l,m)}_g(g_{\rm B}), \nn\\
f_{\rm amp.}^{(l,m)}
&=\mathbb{P}^{(l,m)}_f(f_{\rm amp.}), &
g_{\rm amp.}^{(l,m)}
&=\mathbb{P}^{(l,m)}_g(g_{\rm amp.})\, . 
\end{align}
The functions ${}_{-2}Y_{lm}$ are the spin-weighted spherical harmonics. 

Singular behavior in the forward-scattering limit obscures the decomposition of the effective field theory amplitude into partial-wave components. To understand the issue, consider the nonspinning, point-particle component of the tree-level Compton amplitude, 
\begin{equation}
32\pi G M^2 \left( \frac{\mathcal{N}_0\mathcal{N}_0}{ \, 4\omega^2 x^2} \right), \qquad x^2\equiv {p_1\mdot p_2\over 2\omega^2}=
	\sin^2\left(\frac{\theta}{2}\right)\,.
\end{equation}
This expression diverges as $x \rightarrow 0$ and yields an ill-defined projection of $f_{\rm amp.}(\theta, \phi)$. The problem persists at higher post-Minkowskian orders, where the loop amplitude contains terms with leading singularities of the form $\frac{\log(x)}{x^2}$, $\frac{\log^2(x)}{x^2}$, and so on. These terms arise from the $1/r$ component of the particle's gravitational field and are naturally regularized by the compact expression that captures these Newtonian effects~\cite{Bjerrum-Bohr:2026glv,Dolan:2007ut},
\begin{align}\label{eq:Newt}
32\pi G M^2 \frac{\Gamma(1 - i\bar{\epsilon})}{\Gamma(1 + i\bar{\epsilon})}\left( \frac{\mathcal{N}_0\mathcal{N}_0}{4\omega^2x^{2(1-i\bar{\epsilon})}} \right)\,,
\end{align}
where $\bar{\epsilon}\equiv2GM\omega$. The Newtonian terms in the amplitude match eq.~\eqref{eq:Newt} order by order in the post-Minkowskian expansion. In the form $\frac{1}{x^{2(1-i \bar \epsilon)}}$, the forward-limit singularity is sufficiently softened, and the projection onto $f_{\rm amp.}^{(l,m)}$ can proceed.

All other terms in the Kerr-Compton amplitude have weaker forward-limit singularities, and their projection integrals are well-defined order by order in the post-Minkowskian expansion. In particular, for the tree-level spin-dependent contributions, which are the focus of this paper, any potential forward divergence can only come from the term containing the massless pole, as illustrated below,
 \begin{align}
 \frac{\mathcal{N}_0\mathcal{N}_a-\mathcal{N}_0\mathcal{N}_0}{4\omega^2x^{2}}\,\xrightarrow{x\to 0}  \begin{tikzpicture}[baseline={([yshift=-0.8ex]current bounding box.center)}]\tikzstyle{every node}=[font=\small]    
   \begin{feynman}
    \vertex (a)[dot]{};
     \vertex [above=1.0cm of a](b)[dot]{};
     \vertex [above=0.5cm of a](bc)[]{};
     \vertex [left=0.5cm of bc](bcl)[]{};
     \vertex [right=0.5cm of bc](bcr)[]{};
     \vertex [left=1.1cm of b](c);
     \vertex [left=0.22cm of b](c23);
     \vertex [above=0.13cm of c23](v23)[]{};
    \vertex [above=.4cm of c](j1){$~$};
    \vertex [right=.9cm of j1](j2){$~$};
    \vertex [right=1.3cm of j2](j3){$~$};
    \vertex [left=1.1cm of a](p1)[]{$~$};
    \vertex [right=1.1cm of a](p2)[]{$~$};
   	 \diagram*{(a) -- [photon, ultra thick] (b),(b) -- [photon,ultra thick] (j1),(b)--[photon,ultra thick](j3),(bcl)--[dashed,red,thick](bcr),(p1) -- [fermion, thick] (a)-- [fermion, thick] (p2)};
    \end{feynman}  
  \end{tikzpicture} .
 \end{align}
The massless-cut contribution can be constructed by gluing an on-shell three-point graviton amplitude to the three-point Kerr amplitude~\cite{Arkani-Hamed:2017jhn,Vines:2017hyw,Guevara:2018wpp,Chung:2018kqs,Chen:2022kpm},
 \begin{equation}
 (mv\mdot \veps)\Big(m v\mdot \veps (\cosh \left(q\mdot a\right)-1)+i (q\mdot S\mdot \veps)\frac{ \sinh \left(q\mdot a\right) }{q\mdot a}\Big)\,,
 \end{equation}
where $q^\mu=p^\mu_2-p^\mu_1$. In the forward limit, the spin-dependent terms in the numerator are always multiplied by a factor of $|a||q|=2|a|x\omega$, which softens the singularity.

Employing eq.~\eqref{eq:Newt},
we arrive at the following organization of the Kerr-Compton amplitude to all orders in the post-Minkowskian expansion,
 \begin{align}\label{eq:all-loop Compton}
        \mathcal{M}_{\text{fin.}}(2e^{-\gamma_{\rm E}}\omega, \varepsilon_1, \varepsilon_2^*) &= 32\pi G M^2 \frac{\Gamma(1 - i\bar{\epsilon})}{\Gamma(1 + i\bar{\epsilon})}\Big( \frac{\mathcal{N}_0\mathcal{N}_0}{4\omega^2x^{2(1-i\bar{\epsilon})}} \Big)\nn\\
        &+32\pi G M^2\Big(\frac{\mathcal{N}_0\mathcal{N}_a-\mathcal{N}_0\mathcal{N}_0}{4\omega^2x^{2}}-{1\over 4M^2\omega^2}\mathcal{N}_r+\CN_{\rm c}\Big)\nn\\
        &+ \mathcal{M}_{\text{fin.,rm}}^{\text{Loop}}(2e^{-\gamma_{\rm E}}\omega, \varepsilon_1, \varepsilon_2^*) \,.
    \end{align}
Here the subscript ``rm'' denotes the remaining loop contributions not captured by the resummed leading forward-divergent terms in the first line. In this form, we can readily project onto partial-wave components and apply information from black hole perturbation theory to constrain $\CN_{\rm c}$. The explicit conditions,
\begin{align}\label{eq:compAllComponent}
f_{\rm amp.}^{(l,m)}
&=
f_{\rm B}^{(l,m)},
&
g_{\rm amp.}^{(l,m)}
&=
g_{\rm B}^{(l,m)},
\end{align}
form the basis of the matching process.
\section{Procedure for Extracting the Tree-level Amplitude}\label{sec:TreeKerr}
In this section, we propose an iterative procedure for determining contact terms based on the constraints in eq.~\eqref{eq:compAllComponent}. We begin by constructing an ansatz for $\CN_{\rm c}$ that contains functions of gauge-invariant scalar products among the physical vectors $a^\mu$, $v^\mu$, $p^\mu_1$, and $p^\mu_2$ and the field-strength tensors $F^{\mu\nu}_1$ and $F^{*\mu\nu}_2$. The coefficients of various terms in the ansatz are then fixed by imposing the matching constraints. For a gravitational amplitude, the functions should be linear in $F^{\mu_1\nu_1}_1 F^{\mu_2\nu_2}_1  F^{*\mu_3\nu_3}_2  F^{*\mu_4\nu_4}_2$ for even spin orders and in  $\widetilde F^{\mu_1\nu_1}_1 F^{\mu_2\nu_2}_1  F^{*\mu_3\nu_3}_2  F^{*\mu_4\nu_4}_2$ (or $ F^{\mu_1\nu_1}_1 F^{\mu_2\nu_2}_1  \widetilde F^{*\mu_3\nu_3}_2  F^{*\mu_4\nu_4}_2$) for odd spin orders. To compare with \cite{Bjerrum-Bohr:2023iey} and build upon previously identified terms therein, we split the complete set of possible contact terms into two parts,
\begin{align}
\CN_{\rm c}=\CN_{\rmS,\rm c}+\CN_{\rmA,\rm c}.
\end{align}
Terms in $\CN_{\rmS,\rm c}$ have the following form,
\begin{align}
\label{eq:type1ans}
 \mathcal{G}(a\mdot p_1, a\mdot p_2) \, U(v,a,p_1,p_2,\widetilde F_1, F_1, \widetilde F_2^*, F_2^*)\, ,
\end{align}
where $U$ is a manifestly gauge-invariant function constructed from the dot products of the appropriate field-strength combinations with physical vectors, and thus contains no explicit $a\mdot p_1$, $ a\mdot p_2$, $ v\mdot p_1$ or $ v\mdot p_2$. We keep $U$ dimensionless by including the appropriate power of the spin length $|a|$. \black The function $\mathcal{G}$ is composed of the entire functions  
$G_1$ and $G_2$ defined in Section~\ref{sec:ampstructure}. All other terms appearing in $\CN_{\rm c}$ belong to $\CN_{\rmA,\rm c}$. (We remark that the separation into $\CN_{\rmS,\rm c}$ and $\CN_{\rmA,\rm c}$ is not strictly necessary and is introduced only for computational convenience and easy comparison with previous results.) 

In~\cite{Bjerrum-Bohr:2023iey}, $\CN_{\rmS,\rm c}$ was fixed using the spin-dependent far-zone data at fourth and fifth orders in spin obtained in~\cite{Bautista:2022wjf} from the Teukolsky equation. Starting from the same ansatz as in~\cite{Bjerrum-Bohr:2023iey}, we now reproduce that result using eq.~\eqref{eq:compAllComponent} at linear order in $\epsbar$ for all modes up to fourth order in spin, and the $l=3$ modes at fifth order in spin. Imposing eq.~\eqref{eq:compAllComponent} at $\epsbar$-linear and quartic order in spin uniquely fixes all even-spin contributions to $\CN_{\rmS,\rm c}$ ({\it i.e.}, those at orders $S^4,S^6,S^8,\ldots$) to 
\begin{align}
&\frac{(\partial_{a\mdot p_1}+\partial_{a\mdot p_2})^2 G_1(a\mdot p_1)G_1(-a\mdot p_2)}{2!}\\ \nn
 &\times\bigg( \frac{a\mdot a}{2}\big( a\mdot F_1\mdot F_2^*\mdot v \, a\mdot F_2^*\mdot F_1\mdot v-a\mdot F_1\mdot F_2^*\mdot a\, v\mdot F_1\mdot F_2^*\mdot v \big)-(a\mdot F_1\mdot v)(a\mdot F_2^*\mdot v)(a\mdot F_1\mdot F_2^*\mdot a)  \bigg),
\end{align}
while the odd-spin contributions ({\it i.e.} $S^5,S^7,S^9,\cdots$), 
 \begin{align}\label{eq:oddspin}
   \frac{i(\partial_{a\mdot p_1}+\partial_{a\mdot p_2})^2G_2(a\mdot p_1,-a\mdot p_2)}{2!}&\Big( c_1 \big((a\mdot F_1\mdot F_2^*\mdot a) (a\mdot F_2^*\mdot v) (a\mdot \widetilde F_1\mdot v)-(1\leftrightarrow 2)\big)\nn\\
    &+c_2 \big(a^2 (a\mdot F_2^*\mdot F_1\mdot  v) (a\mdot \widetilde F_1\mdot F_2^*\mdot v)-(1\leftrightarrow 2)\big)\Big)\,,
\end{align} 
have two free parameters $c_1$ and $c_2$ that can be determined by the constraint equation~\eqref{eq:compAllComponent} on the $(l=3,m=3)$ mode at the fifth order in spin,
\begin{equation}\label{eq:mode33}
    \begin{aligned}
{1\over 8\pi M}&\Big(\mathbb{P}^{(3,3)}_f\Big[{\mathcal{M}_{\text{fin.}}(2e^{-\gamma_{\rm E}}\omega, \varepsilon_1^+, \varepsilon_2^{+*})}\Big]\Big|_{\epsbar |a|^5}\Big)=f_{\rm B}^{(3,3)}\Big|_{\epsbar |a|^5}\\
&={\sqrt{\frac{\pi }{42}}\epsbar |a|^5 \omega ^4\over 5760}  \sin (\gamma ) \big(127 \cos (4 \gamma )-44 \cos (2
   \gamma )+45\big)\, .
    \end{aligned}
\end{equation}
The notation $\Big|_{\epsbar^j |a|^n}$ denotes the extraction of the coefficient of the $\epsbar^j |a|^n$ term. By substituting eqs. \eqref{eq:oddspin} and \eqref{eq:cood} into eq. \eqref{eq:mode33}, we get the following equation,
\begin{equation}
    \begin{aligned}
\frac{1}{5}|a|^5 \sqrt{\frac{2 \pi }{21}} \varepsilon  \omega ^4 \sin (\gamma ) \cos ^2(\gamma ) \left(\left(2 c_1+1\right) \sin ^2(\gamma )-4 c_2\right) =0\,.
    \end{aligned}
\end{equation}
Requiring that this equation is satisfied for all values of $\gamma$ uniquely fixes $c_1=-{1\over 2}$ and $c_2=0$. Putting everything together, we obtain
\begin{align}
\label{eq:type1}
 &\npre_{\rmS,\rm c}=\frac{(\partial_{a\mdot p_1}+\partial_{a\mdot p_2})^2 G_1(a\mdot p_1)G_1(-a\mdot p_2)}{2!}\\
 &\times\bigg( \frac{a\mdot a}{2}\big( a\mdot F_1\mdot F_2^*\mdot v\, a\mdot F_2^*\mdot F_1\mdot v-a\mdot F_1\mdot F_2^*\mdot a\, v\mdot F_1\mdot F_2^*\mdot v \big)-(a\mdot F_1\mdot v)(a\mdot F_2^*\mdot v)(a\mdot F_1\mdot F_2^*\mdot a)  \bigg)\nn\\
 &-\frac{i(\partial_{a\mdot p_1}+\partial_{a\mdot p_2})^2G_2(a\mdot p_1,-a\mdot p_2)}{2!}\times\frac{a\mdot F_1\mdot F_2^*\mdot a}{2}\bigg((a\mdot F_2^*\mdot v)(a\mdot \widetilde F_1\mdot v)-(a\mdot F_1\mdot v)(a\mdot \widetilde F_2^*\mdot v)\bigg) \, .\nn
\end{align}

Although our input is limited to quartic and fifth orders in spin, the resulting amplitude, including the contact terms in eq.~\eqref{eq:type1}, shows agreement with black hole perturbation theory through linear order in $\epsbar$ and well beyond these spin orders. Inclusion of $\CN_{\rmS,\rm c}$ alone reproduces all partial-wave modes through quartic order in spin, with no contribution from $\CN_{\rmA,\rm c}$ required in the tree-level amplitude. At fifth, sixth, and seventh orders in spin, the $\CN_{\rmS,\rm c}$ terms continue to agree with black hole perturbation theory for modes with $l\geq 3$, $l\geq 4$, and $l\geq 5$, respectively. The remaining contributions, which must therefore be supplied by $\CN_{\rmA,\rm c}$, are restricted to $l=2$, $l=2,3$, and $l=2,3,4$, respectively, as summarized in Table~\ref{tab:fitType1}. We have verified this pattern for all modes with $l\leq 6$ through sixth order and $l\leq 7$ at seventh order, and expect it to persist at higher orders in spin. Further support comes from~\cite{Chen:2024mmm}, which showed that the amplitude containing only $\CN_{\rmS,\rm c}$ contact terms agrees with the corresponding contribution in~\cite{Bautista:2023sdf} through eighth order in spin and with those in~\cite{Cangemi:2023bpe} to all orders in spin. Together, these results strongly suggest that the $\CN_{\rmS,\rm c}$ sector of the Kerr amplitude remains valid to arbitrary order in spin.
\begin{table}[h]
\centering
\begin{tabular}{c|c|c|c|c}
  & $S^{j\leq 4}$ &$S^5$& $S^6$ & $S^7$ \\ \hline
  $\CN_{\rmS,\rm c}$  match  & all &$l\geq 3$& $l\geq 4$ &  $l\geq 5$ \\ \hline
 $\CN_{\rmA,\rm c}$ needed  &none&$l=2$& $l=2,3$ & $l=2,3,4$
    \end{tabular}
\caption{The correspondence between the amplitude containing $\CN_{\rmS,\rm c}$ and results in black hole perturbation theory in a partial-wave basis at various orders in spin.}
\label{tab:fitType1}
\end{table}

The task that remains is to identify the required $\CN_{\rmA,\rm c}$ contact terms that capture the additional finite-size effects. Based on the separation into $\CN_{\rmS,\rm c}$ and $\CN_{\rmA,\rm c}$, we rewrite the finite all-loop amplitude in eq.~\eqref{eq:all-loop Compton} as
\begin{align}\label{eq:all-loop Compton2}
        \mathcal{M}_{\text{fin.}}(2e^{-\gamma_{\rm E}}\omega, \varepsilon_1, \varepsilon_2^*) &= \mathcal{M}_{\rmS,\text{fin.}}(2e^{-\gamma_{\rm E}}\omega, \varepsilon_1, \varepsilon_2^*) +\mathcal{M}_{\rmA,\text{fin.}}(2e^{-\gamma_{\rm E}}\omega, \varepsilon_1, \varepsilon_2^*) \, ,
        \end{align}
where the first term is the amplitude without $\CN_{\rmA,\rm c}$,
        \begin{align}\label{eq:SPAmp}
        \mathcal{M}_{\rmS,\text{fin.}}(2e^{-\gamma_{\rm E}}\omega, \varepsilon_1, \varepsilon_2^*)=&
         \ 32\pi G M^2 \frac{\Gamma(1 - i\bar{\epsilon})}{\Gamma(1 + i\bar{\epsilon})}\Big( \frac{\mathcal{N}_0\mathcal{N}_0}{4\omega^2x^{2(1-i\bar{\epsilon})}} \Big)\nn\\
        &+32\pi G M^2\Big(\frac{\mathcal{N}_0\mathcal{N}_a-\mathcal{N}_0\mathcal{N}_0}{4\omega^2x^{2}}-{1\over 4M^2\omega^2}\mathcal{N}_r+\CN^{\rmS}_{\rm c}\Big)\nn\\
        &+ \mathcal{M}_{\rmS,\text{fin.,rm}}^{\text{Loop}}(2e^{-\gamma_{\rm E}}\omega, \varepsilon_1, \varepsilon_2^*) \,,
    \end{align}
and the second term contains only the contributions from $\CN_{\rmA,\rm c}$,
\begin{align}\label{eq:AmpE}
\mathcal{M}_{\rmA,\text{fin.}}(2e^{-\gamma_{\rm E}}\omega, \varepsilon_1, \varepsilon_2^*)=\mathcal{M}_{\rmA,\rm c}(\varepsilon_1, \varepsilon_2^*)+ \mathcal{M}_{\rmA,\text{fin.}}^{\text{Loop}}(2e^{-\gamma_{\rm E}}\omega, \varepsilon_1, \varepsilon_2^*)\,.
\end{align}
Here, $\mathcal{M}_{\rmA,\rm c}(\varepsilon_1, \varepsilon_2^*)
\equiv
32\pi G M^2\CN_{\rmA,\rm c}$ and the associated loop contributions $\mathcal{M}_{\rmA,\text{fin.}}^{\text{Loop}}$ are constructed via unitarity or velocity cuts from tree-level building blocks that contain at least one $\CN_{\rmA,\rm c}$ insertion.

Consider an expansion of the tree amplitude in eq.~\eqref{eq:AmpE} in orders of spin and $\epsbar$,
\begin{align}\label{eq:expTree2}
\mathcal{M}_{\rmA,\rm c}( \varepsilon_1, \varepsilon_2^*)=\sum_{n,j}\mathcal{M}^{(j,n)}_{\rmA,\rm c}\equiv \sum_{n,j}\begin{tikzpicture}[scale=1.5, transform shape, baseline=0.1cm]
\tikzstyle{every node}=[font=\small]
\begin{feynman}
    \vertex (a) {};
    \vertex [right=0.6cm of a] (f2) [myblob2] {$\,\mathbf{j}\,$};
     \vertex [above=0.4cm of f2]{$n$};
    \vertex [right=0.6cm of f2] (c) {};
    \vertex [above=0.5cm of f2] (gm) {};
    \vertex [left=0.5cm of gm] (g2) {};
    \vertex [right=0.5cm of gm] (g20) {};
    \diagram* {
        (a) -- [fermion,thick] (f2) -- [fermion,thick] (c),
        (g2) -- [photon, ultra thick] (f2),
        (f2) -- [photon, ultra thick] (g20)
    };
\end{feynman}
\end{tikzpicture},
\end{align}
where $n$ denotes spin order, and $j$ denotes the order in $\epsbar$. The post-Minkowskian order of each term in the sum is $n+j$. Based on the comparison with black hole perturbation theory, we expect possible terms in $\CN_{\rmA,\rm c}$ to be organized as shown in Table~\ref{tab:oderingContact} with contributions starting at fifth post-Minkowskian order.
\begin{table} 
    \begin{tabular}{c|c|c|c|c|c|c|c|c|c}
         & $S^0$ & $S^1$ &  $S^2$& $S^3$ & $S^4$ & $S^5$ & $S^6$ & $S^7$&$\cdots$ \\\hline
        {1PM} & 
$\times$ &  &  &  & &  &  & &\\ \hline
      {\rm 2PM}   & $\times$ & $\times$ &  &  &  &  &  & &\\ \hline
     {\rm 3PM}    & $\times$ & $\times$ & $\times$ &  &  &  &  & &\\ \hline
    {\rm 4PM}     & $\times$ & $\times$ &  $\times$ & $\times$&  &  &  & &\\ \hline
      \begin{tikzpicture}[scale=1.0,baseline=-0.1cm]
\begin{feynman}
 \vertex (v3) at (0.5,0.25) {\rm 5PM};
 \draw[white, line width=1pt]
    (1,0) -- (1,0.5);
\draw[white, line width=1pt]
    (0,0) -- (1,0);
\draw[white, line width=1pt]
    (0,0) -- (0,0.5);
\draw[white, line width=1pt]
    (0,0.5) -- (1,0.5);
\end{feynman}
\end{tikzpicture}  & \begin{tikzpicture}[scale=1.0,baseline=0cm]
\begin{feynman}
 \vertex (v3) at (0.5,0.25) []{  $\,\times\,$};
 \draw[white, line width=1pt]
    (1,0) -- (1,0.5);
\draw[white, line width=1pt]
    (0,0) -- (1,0);
\draw[white, line width=1pt]
    (0,0) -- (0,0.5);
\draw[white, line width=1pt]
    (0,0.5) -- (1,0.5);
\end{feynman}
\end{tikzpicture} & \begin{tikzpicture}[scale=1.0,baseline=0cm]
\begin{feynman}
 \vertex (v3) at (0.5,0.25) [myblob2]{  $\,\mathbf{4}\,$};
\end{feynman}
\draw[gray, line width=1pt]
    (1,0) -- (1,0.5);
\draw[gray, line width=1pt]
    (0,0) -- (1,0);
\draw[gray, line width=1pt]
    (0,0) -- (0,0.5);
\draw[gray, line width=1pt]
    (0,0.5) -- (1,0.5);
\end{tikzpicture} & \begin{tikzpicture}[scale=1.0,baseline=0cm]
\begin{feynman}
 \vertex (v3) at (0.5,0.25) [myblob2]{  $\,\mathbf{3}\,$};
 \draw[gray, line width=1pt]
    (1,0) -- (1,0.5);
\draw[gray, line width=1pt]
    (0,0) -- (1,0);
\draw[gray, line width=1pt]
    (0,0) -- (0,0.5);
\draw[gray, line width=1pt]
    (0,0.5) -- (1,0.5);
\end{feynman}
\end{tikzpicture} & \begin{tikzpicture}[scale=1.0,baseline=0cm]
\begin{feynman}
 \vertex (v3) at (0.5,0.25) [myblob2]{  $\,\mathbf{2}\,$};
 \draw[gray, line width=1pt]
    (1,0) -- (1,0.5);
\draw[gray, line width=1pt]
    (0,0) -- (1,0);
\draw[gray, line width=1pt]
    (0,0) -- (0,0.5);
\draw[gray, line width=1pt]
    (0,0.5) -- (1,0.5);
\end{feynman}
\end{tikzpicture} &  \begin{tikzpicture}[scale=1.0,baseline=0cm]
\begin{feynman}
 \vertex (v3) at (0.5,0.25) {$\times$};
 \draw[white, line width=1pt]
    (1,0) -- (1,0.5);
\draw[white, line width=1pt]
    (0,0) -- (1,0);
\draw[white, line width=1pt]
    (0,0) -- (0,0.5);
\draw[white, line width=1pt]
    (0,0.5) -- (1,0.5);
\end{feynman}
\end{tikzpicture} &  &  & \\ \hline
      \begin{tikzpicture}[scale=1.0,baseline=-0.1cm]
\begin{feynman}
 \vertex (v3) at (0.5,0.25) {\rm 6PM};
 \draw[white, line width=1pt]
    (1,0) -- (1,0.5);
\draw[white, line width=1pt]
    (0,0) -- (1,0);
\draw[white, line width=1pt]
    (0,0) -- (0,0.5);
\draw[white, line width=1pt]
    (0,0.5) -- (1,0.5);
\end{feynman}
\end{tikzpicture}    & \begin{tikzpicture}[scale=1.0,baseline=0cm]
\begin{feynman}
 \vertex (v3) at (0.5,0.25) [myblob2]{  $\,\mathbf{6}\,$};
 \draw[blue, line width=1pt]
    (1,0) -- (1,0.5);
\draw[blue, line width=1pt]
    (0,0) -- (1,0);
\draw[blue, line width=1pt]
    (0,0) -- (0,0.5);
\draw[blue, line width=1pt]
    (0,0.5) -- (1,0.5);
\end{feynman}
\end{tikzpicture} & \begin{tikzpicture}[scale=1.0,baseline=0cm]
\begin{feynman}
 \vertex (v3) at (0.5,0.25) [myblob2]{  $\,\mathbf{5}\,$};
 \draw[gray, line width=1pt]
    (1,0) -- (1,0.5);
\draw[gray, line width=1pt]
    (0,0) -- (1,0);
\draw[gray, line width=1pt]
    (0,0) -- (0,0.5);
\draw[gray, line width=1pt]
    (0,0.5) -- (1,0.5);
\end{feynman}
\end{tikzpicture} & \begin{tikzpicture}[scale=1.0,baseline=0cm]
\begin{feynman}
 \vertex (v3) at (0.5,0.25) [myblob2]{  $\,\mathbf{4}\,$};
 \draw[gray, line width=1pt]
    (1,0) -- (1,0.5);
\draw[gray, line width=1pt]
    (0,0) -- (1,0);
\draw[gray, line width=1pt]
    (0,0) -- (0,0.5);
\draw[gray, line width=1pt]
    (0,0.5) -- (1,0.5);
\end{feynman}
\end{tikzpicture} & \begin{tikzpicture}[scale=1.0,baseline=0cm]
\begin{feynman}
 \vertex (v3) at (0.5,0.25) [myblob2]{  $\,\mathbf{3}\,$};
 \draw[gray, line width=1pt]
    (1,0) -- (1,0.5);
\draw[gray, line width=1pt]
    (0,0) -- (1,0);
\draw[gray, line width=1pt]
    (0,0) -- (0,0.5);
\draw[gray, line width=1pt]
    (0,0.5) -- (1,0.5);
\end{feynman}
\end{tikzpicture} &\begin{tikzpicture}[scale=1.0,baseline=0cm]
\begin{feynman}
 \vertex (v3) at (0.5,0.25) [myblob2]{  $\,\mathbf{2}\,$};
 \draw[gray, line width=1pt]
    (1,0) -- (1,0.5);
\draw[gray, line width=1pt]
    (0,0) -- (1,0);
\draw[gray, line width=1pt]
    (0,0) -- (0,0.5);
\draw[gray, line width=1pt]
    (0,0.5) -- (1,0.5);
\end{feynman}
\end{tikzpicture}  & \begin{tikzpicture}[scale=1.0,baseline=0cm]
\begin{feynman}
 \vertex (v3) at (0.5,0.25) [myblob2]{  $\,\mathbf{1}\,$};
 \draw[red, line width=1pt]
    (1,0) -- (1,0.5);
\draw[red, line width=1pt]
    (0,0) -- (1,0);
\draw[red, line width=1pt]
    (0,0) -- (0,0.5);
\draw[red, line width=1pt]
    (0,0.5) -- (1,0.5);
\end{feynman}
\end{tikzpicture} & \begin{tikzpicture}[scale=1.0,baseline=0cm]
\begin{feynman}
 \vertex (v3) at (0.5,0.25) [myblob2]{  $\,\mathbf{0}\,$};
 \draw[red, line width=1pt]
    (1,0) -- (1,0.5);
\draw[red, line width=1pt]
    (0,0) -- (1,0);
\draw[red, line width=1pt]
    (0,0) -- (0,0.5);
\draw[red, line width=1pt]
    (0,0.5) -- (1,0.5);
\end{feynman}
\end{tikzpicture} & \begin{tikzpicture}[scale=1.0,baseline=0.cm]
\begin{feynman}
 \vertex (v3) at (0.5,0.25) [myblob2]{  -$\mathbf{1}$};
 \draw[red, line width=1pt]
    (1,0) -- (1,0.5);
\draw[red, line width=1pt]
    (0,0) -- (1,0);
\draw[red, line width=1pt]
    (0,0) -- (0,0.5);
\draw[red, line width=1pt]
    (0,0.5) -- (1,0.5);
\end{feynman}
\end{tikzpicture}& \color{red}{\shortstack{low\\$\epsbar$-order}}\\ \hline
     \begin{tikzpicture}[scale=1.0,baseline=-0.1cm]
\begin{feynman}
 \vertex (v3) at (0.5,0.25) {\rm 7PM};
 \draw[white, line width=1pt]
    (1,0) -- (1,0.5);
\draw[white, line width=1pt]
    (0,0) -- (1,0);
\draw[white, line width=1pt]
    (0,0) -- (0,0.5);
\draw[white, line width=1pt]
    (0,0.5) -- (1,0.5);
\end{feynman}
\end{tikzpicture}   & \begin{tikzpicture}[scale=1.0,baseline=0cm]
\begin{feynman}
 \vertex (v3) at (0.5,0.25) [myblob2]{  $\,\mathbf{7}\,$};
 \draw[blue, line width=1pt]
    (1,0) -- (1,0.5);
\draw[blue, line width=1pt]
    (0,0) -- (1,0);
\draw[blue, line width=1pt]
    (0,0) -- (0,0.5);
\draw[blue, line width=1pt]
    (0,0.5) -- (1,0.5);
\end{feynman}
\end{tikzpicture} & \begin{tikzpicture}[scale=1.0,baseline=0cm]
\begin{feynman}
 \vertex (v3) at (0.5,0.25) [myblob2]{  $\,\mathbf{6}\,$};
 \draw[gray, line width=1pt]
    (1,0) -- (1,0.5);
\draw[gray, line width=1pt]
    (0,0) -- (1,0);
\draw[gray, line width=1pt]
    (0,0) -- (0,0.5);
\draw[gray, line width=1pt]
    (0,0.5) -- (1,0.5);
\end{feynman}
\end{tikzpicture} & \begin{tikzpicture}[scale=1.0,baseline=0cm]
\begin{feynman}
 \vertex (v3) at (0.5,0.25) [myblob2]{  $\,\mathbf{5}\,$};
 \draw[gray, line width=1pt]
    (1,0) -- (1,0.5);
\draw[gray, line width=1pt]
    (0,0) -- (1,0);
\draw[gray, line width=1pt]
    (0,0) -- (0,0.5);
\draw[gray, line width=1pt]
    (0,0.5) -- (1,0.5);
\end{feynman}
\end{tikzpicture} & \begin{tikzpicture}[scale=1.0,baseline=0cm]
\begin{feynman}
 \vertex (v3) at (0.5,0.25) [myblob2]{  $\,\mathbf{4}\,$};
 \draw[gray, line width=1pt]
    (1,0) -- (1,0.5);
\draw[gray, line width=1pt]
    (0,0) -- (1,0);
\draw[gray, line width=1pt]
    (0,0) -- (0,0.5);
\draw[gray, line width=1pt]
    (0,0.5) -- (1,0.5);
\end{feynman}
\end{tikzpicture} & \begin{tikzpicture}[scale=1.0,baseline=0cm]
\begin{feynman}
 \vertex (v3) at (0.5,0.25) [myblob2]{  $\,\mathbf{3}\,$};
 \draw[gray, line width=1pt]
    (1,0) -- (1,0.5);
\draw[gray, line width=1pt]
    (0,0) -- (1,0);
\draw[gray, line width=1pt]
    (0,0) -- (0,0.5);
\draw[gray, line width=1pt]
    (0,0.5) -- (1,0.5);
\end{feynman}
\end{tikzpicture} & \begin{tikzpicture}[scale=1.0,baseline=0cm]
\begin{feynman}
 \vertex (v3) at (0.5,0.25) [myblob2]{  $\,\mathbf{2}\,$};
 \draw[gray, line width=1pt]
    (1,0) -- (1,0.5);
\draw[gray, line width=1pt]
    (0,0) -- (1,0);
\draw[gray, line width=1pt]
    (0,0) -- (0,0.5);
\draw[gray, line width=1pt]
    (0,0.5) -- (1,0.5);
\end{feynman}
\end{tikzpicture} & \begin{tikzpicture}[scale=1.0,baseline=0cm]
\begin{feynman}
 \vertex (v3) at (0.5,0.25) [myblob2]{  $\,\mathbf{1}\,$};
 \draw[red, line width=1pt]
    (1,0) -- (1,0.5);
\draw[red, line width=1pt]
    (0,0) -- (1,0);
\draw[red, line width=1pt]
    (0,0) -- (0,0.5);
\draw[red, line width=1pt]
    (0,0.5) -- (1,0.5);
\end{feynman}
\end{tikzpicture} & \begin{tikzpicture}[scale=1.0,baseline=0cm]
\begin{feynman}
 \vertex (v3) at (0.5,0.25) [myblob2]{  $\,\mathbf{0}\,$};
 \draw[red, line width=1pt]
    (1,0) -- (1,0.5);
\draw[red, line width=1pt]
    (0,0) -- (1,0);
\draw[red, line width=1pt]
    (0,0) -- (0,0.5);
\draw[red, line width=1pt]
    (0,0.5) -- (1,0.5);
\end{feynman}
\end{tikzpicture}&\\ \hline
     \begin{tikzpicture}[scale=1.0,baseline=-0.1cm]
\begin{feynman}
 \vertex (v3) at (0.5,0.25) {\rm 8PM};
 \draw[white, line width=1pt]
    (1,0) -- (1,0.5);
\draw[white, line width=1pt]
    (0,0) -- (1,0);
\draw[white, line width=1pt]
    (0,0) -- (0,0.5);
\draw[white, line width=1pt]
    (0,0.5) -- (1,0.5);
\end{feynman}
\end{tikzpicture}     & \begin{tikzpicture}[scale=1.0,baseline=0cm]
\begin{feynman}
 \vertex (v3) at (0.5,0.25) [myblob2]{  $\,\mathbf{8}\,$};
 \draw[blue, line width=1pt]
    (1,0) -- (1,0.5);
\draw[blue, line width=1pt]
    (0,0) -- (1,0);
\draw[blue, line width=1pt]
    (0,0) -- (0,0.5);
\draw[blue, line width=1pt]
    (0,0.5) -- (1,0.5);
\end{feynman}
\end{tikzpicture} & \begin{tikzpicture}[scale=1.0,baseline=0cm]
\begin{feynman}
 \vertex (v3) at (0.5,0.25) [myblob2]{  $\,\mathbf{7}\,$};
 \draw[gray, line width=1pt]
    (1,0) -- (1,0.5);
\draw[gray, line width=1pt]
    (0,0) -- (1,0);
\draw[gray, line width=1pt]
    (0,0) -- (0,0.5);
\draw[gray, line width=1pt]
    (0,0.5) -- (1,0.5);
\end{feynman}
\end{tikzpicture} & \begin{tikzpicture}[scale=1.0,baseline=0cm]
\begin{feynman}
 \vertex (v3) at (0.5,0.25) [myblob2]{  $\,\mathbf{6}\,$};
 \draw[gray, line width=1pt]
    (1,0) -- (1,0.5);
\draw[gray, line width=1pt]
    (0,0) -- (1,0);
\draw[gray, line width=1pt]
    (0,0) -- (0,0.5);
\draw[gray, line width=1pt]
    (0,0.5) -- (1,0.5);
\end{feynman}
\end{tikzpicture} & \begin{tikzpicture}[scale=1.0,baseline=0cm]
\begin{feynman}
 \vertex (v3) at (0.5,0.25) [myblob2]{  $\,\mathbf{5}\,$};
 \draw[gray, line width=1pt]
    (1,0) -- (1,0.5);
\draw[gray, line width=1pt]
    (0,0) -- (1,0);
\draw[gray, line width=1pt]
    (0,0) -- (0,0.5);
\draw[gray, line width=1pt]
    (0,0.5) -- (1,0.5);
\end{feynman}
\end{tikzpicture} & \begin{tikzpicture}[scale=1.0,baseline=0cm]
\begin{feynman}
 \vertex (v3) at (0.5,0.25) [myblob2]{  $\,\mathbf{4}\,$};
 \draw[gray, line width=1pt]
    (1,0) -- (1,0.5);
\draw[gray, line width=1pt]
    (0,0) -- (1,0);
\draw[gray, line width=1pt]
    (0,0) -- (0,0.5);
\draw[gray, line width=1pt]
    (0,0.5) -- (1,0.5);
\end{feynman}
\end{tikzpicture} & \begin{tikzpicture}[scale=1.0,baseline=0cm]
\begin{feynman}
 \vertex (v3) at (0.5,0.25) [myblob2]{  $\,\mathbf{3}\,$};
 \draw[gray, line width=1pt]
    (1,0) -- (1,0.5);
\draw[gray, line width=1pt]
    (0,0) -- (1,0);
\draw[gray, line width=1pt]
    (0,0) -- (0,0.5);
\draw[gray, line width=1pt]
    (0,0.5) -- (1,0.5);
\end{feynman}
\end{tikzpicture} & \begin{tikzpicture}[scale=1.0,baseline=0cm]
\begin{feynman}
 \vertex (v3) at (0.5,0.25) [myblob2]{  $\,\mathbf{2}\,$};
 \draw[gray, line width=1pt]
    (1,0) -- (1,0.5);
\draw[gray, line width=1pt]
    (0,0) -- (1,0);
\draw[gray, line width=1pt]
    (0,0) -- (0,0.5);
\draw[gray, line width=1pt]
    (0,0.5) -- (1,0.5);
\end{feynman}
\end{tikzpicture} & \begin{tikzpicture}[scale=1.0,baseline=0cm]
\begin{feynman}
 \vertex (v3) at (0.5,0.25) [myblob2]{  $\,\mathbf{1}\,$};
 \draw[red, line width=1pt]
    (1,0) -- (1,0.5);
\draw[red, line width=1pt]
    (0,0) -- (1,0);
\draw[red, line width=1pt]
    (0,0) -- (0,0.5);
\draw[red, line width=1pt]
    (0,0.5) -- (1,0.5);
\end{feynman}
\end{tikzpicture} &\\ \hline
\vdots &\color{blue}{tidal} 
    \end{tabular}
\caption{Possible terms in the tree-level amplitude built from $\mathcal{N}_{\rmA,\rm c}$. Each square in the table, labeled by $j$, corresponds to a term in the expansion defined in eq.~\eqref{eq:expTree2} with the columns counting out $n$ and the rows $n+j$. The entries marked in blue represent spin-independent tidal effects, while those in red form the low $\epsbar$-order ($j\leq 1$)  spin sector. The remaining terms are categorized as the high $\epsbar$-order ($j\geq 2$)  spin sector.  In addition, we use ``$\times$" to mark orders to which the amplitude $\mathcal{M}_{\rmS,\text{fin.}}$ contributes. While this should include all orders in the lower triangle including the diagonal, we only display ``$\times$" where it is not covered by the squares coming from the presence of $\mathcal{N}_{\rmA,\rm c}$.}
\label{tab:oderingContact}
\end{table} 
At sixth post-Minkowskian order, terms with negative powers of $\epsbar$ also appear. Their presence can be traced to non-rational functions of $|a|/(GM)$ that arise in the black hole perturbation theory results; an example being
\begin{align}
\psi ^{(0)}\left(-2+\frac{2 i |a|/(GM)}{\sqrt{1-|a|^2/(GM)^2}}\right), \qquad
\psi ^{(0)}(z)\equiv {d\ln\Gamma(z)\over dz} .
\end{align}
The ratio $|a|/(GM)$ appears naturally as a parameter in the Teukolsky equation and shows up in black hole absorption effects, as manifested in the horizon flux computed within black hole perturbation theory~\cite{Tagoshi:1997jy, Mino:1997bx,Sasaki:2003xr}. Table~\ref{tab:oderingContact} implies that the effective range of summation is such that only a finite number of negative $\epsbar$-orders occur at any fixed spin order,
\begin{align}\label{eq:newterms}
\mathcal{M}_{\rmA,\rm c}( \varepsilon_1, \varepsilon_2^*)=\sum_{n=0}^{3} \sum_{j=5-n}^{\infty}\begin{tikzpicture}[scale=1.5, transform shape, baseline=0.1cm]
\tikzstyle{every node}=[font=\small]
\begin{feynman}
    \vertex (a) {};
    \vertex [right=0.6cm of a] (f2) [myblob2] {$\,\mathbf{j}\,$};
     \vertex [above=0.4cm of f2]{$n$};
    \vertex [right=0.6cm of f2] (c) {};
    \vertex [above=0.5cm of f2] (gm) {};
    \vertex [left=0.5cm of gm] (g2) {};
    \vertex [right=0.5cm of gm] (g20) {};
    \diagram* {
        (a) -- [fermion,thick] (f2) -- [fermion,thick] (c),
        (g2) -- [photon, ultra thick] (f2),
        (f2) -- [photon, ultra thick] (g20)
    };
\end{feynman}
\end{tikzpicture}+\sum_{n=4}^{\infty} \sum_{j=6-n}^{\infty}\begin{tikzpicture}[scale=1.5, transform shape, baseline=0.1cm]
\tikzstyle{every node}=[font=\small]
\begin{feynman}
    \vertex (a) {};
    \vertex [right=0.6cm of a] (f2) [myblob2] {$\,\mathbf{j}\,$};
     \vertex [above=0.4cm of f2]{$n$};
    \vertex [right=0.6cm of f2] (c) {};
    \vertex [above=0.5cm of f2] (gm) {};
    \vertex [left=0.5cm of gm] (g2) {};
    \vertex [right=0.5cm of gm] (g20) {};
    \diagram* {
        (a) -- [fermion,thick] (f2) -- [fermion,thick] (c),
        (g2) -- [photon, ultra thick] (f2),
        (f2) -- [photon, ultra thick] (g20)
    };
\end{feynman}
\end{tikzpicture}\, .
\end{align}
This expansion for the tree amplitude is effective when $|a|<GM$.

We can now write the ansatz for the terms in $\CN_{\rmA,\rm c}$,
\begin{align}
{\epsbar^{j_0}}\,\,\mathcal{P}(a\mdot p_1, a\mdot p_2,|a|\omega,\epsbar,p_1\mdot p_2 \, a\mdot a, p_1\mdot p_2 (GM)^2)\, U(v,a,p_1,p_2,\widetilde F_1, F_1, \widetilde F_2^*,F_2^*)\, ,
\end{align}
where $\mathcal{P}$ is a polynomial in the listed arguments with spin order $n_1$, and $U$ denotes the same type of gauge-invariant functions as in eq.~\eqref{eq:type1ans}, with spin order $n_2$, such that $n_1+n_2=n$. The explicit power of $\epsbar$ appearing in the ansatz satisfies $j_0=5-n$ for $n\geq 6$ and $j_0=0$ otherwise. At fixed post-Minkowskian and spin orders, the number of terms in the ansatz is finite, rendering the matching procedure tractable. Since the black hole perturbation theory amplitude and the amplitude in eq.~\eqref{eq:SPAmp} without $\CN_{\rmA,\rm c}$ are both known, their difference, which can be decomposed into helicity-conserving and helicity-reversing components,
\begin{align}
\label{eq:ampdiff}
f_{\rm diff.}^{(l,m)}&\equiv f_{\rm B}^{(l,m)}-{1\over 8\pi M} \mathbb{P}^{(l,m)}_f\Big[\mathcal{M}_{\rmS, \text{fin.}}(2e^{-\gamma_{\rm E}}\omega, \varepsilon_1^+, \varepsilon_2^{+*})\Big]\, \nn, \\
g_{\rm diff.}^{(l,m)}&\equiv g_{\rm B}^{(l,m)}- {1\over 8\pi M}\mathbb{P}^{(l,m)}_g\Big[\mathcal{M}_{\rmS, \text{fin.}}(2e^{-\gamma_{\rm E}}\omega, \varepsilon_1^+, \varepsilon_2^{-*})\Big]\, ,
\end{align}
provides the following constraints on the $\CN_{\rmA,\rm c}$ contributions,
\begin{align}
\label{eq:type2fix}
&\mathbb{P}^{(l,m)}_f\Big[\mathcal{M}_{\rmA,\rm c}( \varepsilon_1^+, \varepsilon_2^{+*})\Big]+\mathbb{P}^{(l,m)}_f\Big[\mathcal{M}_{\rmA,\text{fin.}}^{\text{Loop}}(2e^{-\gamma_{\rm E}}\omega, \varepsilon_1^+, \varepsilon_2^{+*})\Big]=8\pi M f_{\rm diff.}^{(l,m)}\, ,\nn\\
&\mathbb{P}^{(l,m)}_g\Big[\mathcal{M}_{\rmA,\rm c}(\varepsilon_1^+, \varepsilon_2^{-*})\Big]+\mathbb{P}^{(l,m)}_g\Big[\mathcal{M}_{\rmA,\text{fin.}}^{\text{Loop}}(2e^{-\gamma_{\rm E}}\omega, \varepsilon_1^+, \varepsilon_2^{-*})\Big]=8\pi M g_{\rm diff.}^{(l,m)}\, .
\end{align} 
The difference contains no leading forward divergence. This indicates that the projection integrals on the left-hand side of the above equations are well-defined term by term in the spin and $\epsbar$ expansions.

To illustrate the procedure for fixing $\CN_{\rmA,\rm c}$, we introduce additional diagrammatic notation for the expansion in spin and $\epsbar$ of the loop amplitude generated by $\CN_{\rmA,\rm c}$ and the Compton amplitude without $\CN_{\rmA,\rm c}$, respectively,
\begin{align} 
\mathcal{M}_{\rmA, \text{fin.}}^{\text{Loop}}\Big|_{\epsbar^j|a|^n}\equiv\begin{tikzpicture}[scale=1.0, transform shape, baseline=0.1cm]
\begin{feynman}
    \vertex (a) {};
    \vertex [right=1.cm of a] (f2) [GR] {$\,\,\,\mathbf{j}\,\,\,$};
    \vertex [above=0.6cm of f2]{$n$};
    \vertex [right=1.0cm of f2] (c) {};
    \vertex [above=1.cm of f2] (gm) {};
    \vertex [left=0.5cm of gm] (g2) {};
    \vertex [right=0.5cm of gm] (g20) {};
    \diagram* {
        (a) -- [fermion,thick] (f2) -- [fermion,thick] (c),
        (g2) -- [photon, ultra thick] (f2),
        (f2) -- [photon, ultra thick] (g20)
    };
\end{feynman}
\end{tikzpicture} \, ,
&& 
\mathcal{M}_{\rmS,\text{fin.}}\Big|_{\epsbar^j|a|^n}\equiv \begin{tikzpicture}[scale=1.5, transform shape, baseline=0.1cm]
\tikzstyle{every node}=[font=\small]
\begin{feynman}
    \vertex (a) {};
    \vertex [right=0.6cm of a] (f2) [HV2] {$\mathbf{j}$};
    \vertex [above=0.4cm of f2]{$n$};
    \vertex [right=0.6cm of f2] (c) {};
    \vertex [above=0.5cm of f2] (gm) {};
    \vertex [left=0.5cm of gm] (g2) {};
    \vertex [right=0.5cm of gm] (g20) {};
    \diagram* {
        (a) -- [fermion,thick] (f2) -- [fermion,thick] (c),
        (g2) -- [photon, ultra thick] (f2),
        (f2) -- [photon, ultra thick] (g20)
    };
\end{feynman}
\end{tikzpicture} \, .
\end{align}
With this notation, eq.~\eqref{eq:type2fix} can be written as
\begin{align}\label{eq:extractTree}
\mathbb{P}^{(l,m)}_f\Big[\kern-0.6em\begin{tikzpicture}[scale=1.5, transform shape, baseline=0.1cm]
\tikzstyle{every node}=[font=\small]
\begin{feynman}
    \vertex (a) {};
    \vertex [right=0.6cm of a] (f2) [myblob2] {$\,\mathbf{j}\,$};
    \vertex [above=0.3cm of f2]{$n$};
    \vertex [right=0.6cm of f2] (c) {};
    \vertex [above=0.5cm of f2] (gm) {};
    \vertex [left=0.5cm of gm] (g2) {};
    \vertex [right=0.5cm of gm] (g20) {};
    \diagram* {
        (a) -- [fermion,thick] (f2) -- [fermion,thick] (c),
        (g2) -- [photon, ultra thick] (f2),
        (f2) -- [photon, ultra thick] (g20)
    };
\end{feynman}
\end{tikzpicture}\kern-0.6em\Big]&=8\pi M f_{\rm diff.}^{(l,m)}\Big|_{\epsbar^j |a|^n}-\mathbb{P}^{(l,m)}_f\Big[\kern-0.6em\begin{tikzpicture}[scale=1.0, transform shape, baseline=0.1cm]
\begin{feynman}
    \vertex (a) {};
    \vertex [right=1.cm of a] (f2) [GR] {$\,\,\,\mathbf{j}\,\,\,$};
    \vertex [above=0.6cm of f2]{$n$};
    \vertex [right=1.0cm of f2] (c) {};
    \vertex [above=1.cm of f2] (gm) {};
    \vertex [left=0.5cm of gm] (g2) {};
    \vertex [right=0.5cm of gm] (g20) {};
    \diagram* {
        (a) -- [fermion,thick] (f2) -- [fermion,thick] (c),
        (g2) -- [photon, ultra thick] (f2),
        (f2) -- [photon, ultra thick] (g20)
    };
\end{feynman}
\end{tikzpicture}\kern-0.6em\Big] \, ,\nn\\
\mathbb{P}^{(l,m)}_g \Big[\kern-0.6em\begin{tikzpicture}[scale=1.5, transform shape, baseline=0.1cm]
\tikzstyle{every node}=[font=\small]
\begin{feynman}
    \vertex (a) {};
    \vertex [right=0.6cm of a] (f2) [myblob2] {$\,\mathbf{j}\,$};
    \vertex [above=0.3cm of f2]{$n$};
    \vertex [right=0.6cm of f2] (c) {};
    \vertex [above=0.5cm of f2] (gm) {};
    \vertex [left=0.5cm of gm] (g2) {};
    \vertex [right=0.5cm of gm] (g20) {};
    \diagram* {
        (a) -- [fermion,thick] (f2) -- [fermion,thick] (c),
        (g2) -- [photon, ultra thick] (f2),
        (f2) -- [photon, ultra thick] (g20)
    };
\end{feynman}
\end{tikzpicture}\kern-0.6em\Big]&=8\pi M g_{\rm diff.}^{(l,m)}\Big|_{\epsbar^j |a|^n}-\mathbb{P}^{(l,m)}_g\Big[\kern-0.6em\begin{tikzpicture}[scale=1.0, transform shape, baseline=0.1cm]
\begin{feynman}
    \vertex (a) {};
    \vertex [right=1.cm of a] (f2) [GR] {$\,\,\,\mathbf{j}\,\,\,$};
    \vertex [above=0.6cm of f2]{$n$};
    \vertex [right=1.0cm of f2] (c) {};
    \vertex [above=1.cm of f2] (gm) {};
    \vertex [left=0.5cm of gm] (g2) {};
    \vertex [right=0.5cm of gm] (g20) {};
    \diagram* {
        (a) -- [fermion,thick] (f2) -- [fermion,thick] (c),
        (g2) -- [photon, ultra thick] (f2),
        (f2) -- [photon, ultra thick] (g20)
    };
\end{feynman}
\end{tikzpicture}\kern-0.6em\Big] \, .
\end{align}
We provide the relations for determining three terms at the sixth and seventh post-Minkowskian order as examples, 
\begin{align}
\mathbb{P}_f^{(l,m)}\Big[\kern-0.6em\begin{tikzpicture}[scale=1.5, transform shape, baseline=0.1cm]
\tikzstyle{every node}=[font=\small]
\begin{feynman}
    \vertex (a) {};
    \vertex [right=0.6cm of a] (f2) [myblob2] {$\,\mathbf{2}\,$};
    \vertex [above=0.3cm of f2]{\tiny $4$};
    \vertex [right=0.6cm of f2] (c) {};
    \vertex [above=0.5cm of f2] (gm) {};
    \vertex [left=0.5cm of gm] (g2) {};
    \vertex [right=0.5cm of gm] (g20) {};
    \diagram* {
        (a) -- [fermion,thick] (f2) -- [fermion,thick] (c),
        (g2) -- [photon, ultra thick] (f2),
        (f2) -- [photon, ultra thick] (g20)
    };
\end{feynman}
\end{tikzpicture}\kern-0.6em\Big]&=8\pi M f_{\rm diff.}^{(l,m)}\Big|_{\epsbar^{2} |a|^4}  \, , \quad \mathbb{P}_f^{(l,m)}\Big[\kern-0.6em\begin{tikzpicture}[scale=1.5, transform shape, baseline=0.1cm]
\tikzstyle{every node}=[font=\small]
\begin{feynman}
    \vertex (a) {};
    \vertex [right=0.6cm of a] (f2) [myblob2] {$\,\mathbf{6}\,$};
    \vertex [above=0.3cm of f2]{\tiny $0$};
    \vertex [right=0.6cm of f2] (c) {};
    \vertex [above=0.5cm of f2] (gm) {};
    \vertex [left=0.5cm of gm] (g2) {};
    \vertex [right=0.5cm of gm] (g20) {};
    \diagram* {
        (a) -- [fermion,thick] (f2) -- [fermion,thick] (c),
        (g2) -- [photon, ultra thick] (f2),
        (f2) -- [photon, ultra thick] (g20)
    };
\end{feynman}
\end{tikzpicture}\kern-0.6em\Big]=8\pi M f_{\rm diff.}^{(l,m)}\Big|_{\epsbar^{6} |a|^0}\, ,\nn\\
\mathbb{P}_f^{(l,m)}\Big[\kern-0.6em\begin{tikzpicture}[scale=1.5, transform shape, baseline=0.1cm]
\tikzstyle{every node}=[font=\small]
\begin{feynman}
    \vertex (a) {};
    \vertex [right=0.6cm of a] (f2) [myblob2] {$\,\mathbf{7}\,$};
    \vertex [above=0.3cm of f2]{\tiny $0$};
    \vertex [right=0.6cm of f2] (c) {};
    \vertex [above=0.5cm of f2] (gm) {};
    \vertex [left=0.5cm of gm] (g2) {};
    \vertex [right=0.5cm of gm] (g20) {};
    \diagram* {
        (a) -- [fermion,thick] (f2) -- [fermion,thick] (c),
        (g2) -- [photon, ultra thick] (f2),
        (f2) -- [photon, ultra thick] (g20)
    };
\end{feynman}
\end{tikzpicture}\kern-0.6em\Big]&=8\pi M f_{\rm diff.}^{(l,m)}\Big|_{\epsbar^7 |a|^0}- \mathbb{P}_f^{(l,m)}\Big[\kern-0.6em\begin{tikzpicture}[scale=1.5, transform shape, baseline=0.1cm]
\tikzstyle{every node}=[font=\small]
\begin{feynman}
    \vertex (a) {};
    \vertex [right=0.6cm of a] (f2) [myblob2] {$\,\mathbf{6}\,$};
    \vertex [right=0.7cm of f2] (f3) [HV2] {$\,\mathbf{1}\,$};
    \vertex [above=0.3cm of f2]{\tiny $0$};
    \vertex [above=0.3cm of f3]{\tiny $0$};
    \vertex [right=1.3cm of f2] (c) {};
    \vertex [above=0.5cm of f2] (gm) {};
    \vertex [left=0.5cm of gm] (g2) {};
    \vertex [right=1.2cm of gm] (g20) {};
    \vertex [right=0.35cm of f2] (cut1);
    \vertex [below=0.15cm of cut1] (cut1u);
    \vertex [above=0.1cm of cut1] (cut1b);
    \vertex [above=0.4cm of cut1b] (cut1bb);
    \diagram* {
        (a) -- [fermion,thick] (f2) -- [fermion,thick] (f3)-- [fermion,thick] (c),
        (f2)-- [photon,ultra thick,out=35,in=145,looseness=0.5,min distance=0.2cm] (f3),
        (g2) -- [photon, ultra thick] (f2),
        (f3) -- [photon, ultra thick] (g20),
         (cut1u)--[red,thick] (cut1b),
      (cut1b)--[dashed,red,thick] (cut1bb)
    };
\end{feynman}
\end{tikzpicture}\kern-0.6em +\kern-0.6em\begin{tikzpicture}[scale=1.5, transform shape, baseline=0.1cm]
\tikzstyle{every node}=[font=\small]
\begin{feynman}
    \vertex (a) {};
    \vertex [right=0.6cm of a] (f2) [HV2] {$\,\mathbf{1}\,$};
    \vertex [right=0.7cm of f2] (f3) [myblob2] {$\,\mathbf{6}\,$};
    \vertex [above=0.3cm of f2]{\tiny $0$};
    \vertex [above=0.3cm of f3]{\tiny $0$};
    \vertex [right=1.3cm of f2] (c) {};
    \vertex [above=0.5cm of f2] (gm) {};
    \vertex [left=0.5cm of gm] (g2) {};
    \vertex [right=1.2cm of gm] (g20) {};
    \vertex [right=0.35cm of f2] (cut1);
    \vertex [below=0.15cm of cut1] (cut1u);
    \vertex [above=0.1cm of cut1] (cut1b);
    \vertex [above=0.4cm of cut1b] (cut1bb);
    \diagram* {
        (a) -- [fermion,thick] (f2) -- [fermion,thick] (f3)-- [fermion,thick] (c),
        (f2)-- [photon,ultra thick,out=35,in=145,looseness=0.5,min distance=0.2cm] (f3),
        (g2) -- [photon, ultra thick] (f2),
        (f3) -- [photon, ultra thick] (g20),
         (cut1u)--[red,thick] (cut1b),
      (cut1b)--[dashed,red,thick] (cut1bb)
    };
\end{feynman}
\end{tikzpicture}\kern-0.6em\Big].
\end{align}
For the first two examples, there is no $\mathcal{M}_{\rmA, \text{fin.}}^{\text{Loop}}$ contribution that needs to be subtracted. To extract the spinless contact contribution in the third example, we need to consider the loop amplitude constructed from the amplitude at first post-Minkowskian order and the spinless contributions to $\CN_{\rmA,\rm c}$ at sixth post-Minkowskian order. More generally, at a fixed order $\bar{\epsilon}^j |a|^n$, knowing all lower-order terms in $\CN_{\rmA,\rm c}$ fixes the associated loop contributions on the right-hand side of eq.~\eqref{eq:extractTree}. Combining these loop contributions with $f_{\rm diff.}^{(l,m)}$ and $g_{\rm diff.}^{(l,m)}$, calculated according to eq.~\eqref{eq:ampdiff}, uniquely determines the contact terms at the order under consideration. Repeating this procedure order by order provides an iterative construction of the terms in $\CN_{\rmA,\rm c}$.
\section{Determining Low $\epsbar$-order Spin Contact Terms}\label{sec:contactTerms}
The amplitude $\mathcal{M}_{\rmS,\text{fin.}}$ is needed to evaluate $f_{\rm diff.}^{(l,m)}$ and $g_{\rm diff.}^{(l,m)}$ and thereby determine the terms in $\CN_{\rmA,\rm c}$. In general, determining $\mathcal{M}_{\rmS,\text{fin.}}$ for contact terms associated with spin-independent tidal effects or the high $\bar{\epsilon}$-order spin sector requires challenging loop calculations. By contrast, determining the contact terms in the low $\bar{\epsilon}$-order spin sector, shown by the red entries in Table~\ref{tab:oderingContact}, is considerably simpler. In this sector, terms in $\mathcal{M}_{\rmS, \text{fin.}}$ need to be extracted only to linear order in $\bar{\epsilon}$, and no loop calculation is needed to identify the difference between $\mathcal{M}_{\rmS, \text{fin.}}$ and the black hole perturbation theory results. Loops arise solely from the second term in eq.~\eqref{eq:extractTree}. Consequently, determining the contact terms at the $n$-th order in spin requires amplitudes up to $(n-5)$ loops, which greatly improves computational efficiency. For example, the contact contribution at fifth order in spin requires only a tree-level calculation, while the contribution at sixth order in spin requires a one-loop calculation. 
\subsection{At Fifth Order in Spin}
We first present the difference between the scattering functions in black hole perturbation theory and $\mathcal{M}_{\rmS, \text{fin.}}$, in the spherical harmonic basis.
As indicated in Table~\ref{tab:fitType1}, the helicity-conserving components agree for $l\geq 3$ while the helicity-reversing components agree completely, as explicitly verified for $l\leq 6$. Thus, the only nontrivial difference at linear order in $\epsbar$ occurs in the $l=2$ helicity-conserving components,
\begin{align}
f_{\rm diff.}^{(2,-2)}\Big|_{\epsbar |a|^5}&=0,~~~
f_{\rm diff.}^{(2,-1)}\Big|_{\epsbar |a|^5}=4 \sqrt{\frac{\pi }{5}} \epsbar \omega ^4 | a| ^5 \sin ^3(\gamma ) ( \mathsf{b}_1- \mathsf{b}_2),\nn\\
f_{\rm diff.}^{(2,0)}\Big|_{\epsbar |a|^5}&=-4 \sqrt{\frac{6 \pi }{5}} \epsbar \omega ^4 | a| ^5 \sin ^2(\gamma ) \cos (\gamma ) ( \mathsf{b}_1- \mathsf{b}_2),\nn\\
f_{\rm diff.}^{(2,1)}\Big|_{\epsbar |a|^5}&=2 \sqrt{\frac{\pi }{5}} \epsbar \omega ^4 | a| ^5 \sin (\gamma ) (3 \cos (2 \gamma ) ( \mathsf{b}_1- \mathsf{b}_2)+ \mathsf{b}_1-5  \mathsf{b}_2)\,,\nn\\
f_{\rm diff.}^{(2,2)}\Big|_{\epsbar |a|^5}&=-4 \sqrt{\frac{\pi }{5}} \epsbar \omega ^4 | a| ^5 \cos (\gamma ) (\cos (2 \gamma ) ( \mathsf{b}_1- \mathsf{b}_2)- \mathsf{b}_1-3  \mathsf{b}_2)\, ,
\end{align}
where 
\begin{align}
\mathsf{b}_1=\frac{2}{45}\,\zeta(3)-\frac{8}{45}\,\zeta(5)+\frac{1}{18},\quad \mathsf{b}_2=\frac{128}{45}\,\zeta(3)-\frac{128}{45}\,\zeta(5)-\frac{14}{45}.
\end{align}
According to eq.~\eqref{eq:extractTree}, the contact contribution in the low $\epsbar$-order sector at fifth order in spin must thus satisfy 
\begin{align}
\label{eq:s5cons}
\mathbb{P}_f^{(l,m)}\Big[\kern-0.6em\begin{tikzpicture}[scale=1.5, transform shape, baseline=0.1cm]
\tikzstyle{every node}=[font=\small]
\begin{feynman}
    \vertex (a) {};
    \vertex [right=0.6cm of a] (f2) [myblob2] {$\,\mathbf{1}\,$};
    \vertex [above=0.3cm of f2]{\tiny $5$};
    \vertex [right=0.6cm of f2] (c) {};
    \vertex [above=0.5cm of f2] (gm) {};
    \vertex [left=0.5cm of gm] (g2) {};
    \vertex [right=0.5cm of gm] (g20) {};
    \diagram* {
        (a) -- [fermion,thick] (f2) -- [fermion,thick] (c),
        (g2) -- [photon, ultra thick] (f2),
        (f2) -- [photon, ultra thick] (g20)
    };
\end{feynman}
\end{tikzpicture}\kern-0.6em\Big]&=\begin{cases}{8\pi M} f_{\rm diff.}^{(2,m)}\Big|_{\epsbar |a|^5},& l=2\\
0,&l\neq 2
\end{cases}\, , &\mathbb{P}_g^{(l,m)}\Big[\kern-0.6em\begin{tikzpicture}[scale=1.5, transform shape, baseline=0.1cm]
\tikzstyle{every node}=[font=\small]
\begin{feynman}
    \vertex (a) {};
    \vertex [right=0.6cm of a] (f2) [myblob2] {$\,\mathbf{1}\,$};
    \vertex [above=0.3cm of f2]{\tiny $5$};
    \vertex [right=0.6cm of f2] (c) {};
    \vertex [above=0.5cm of f2] (gm) {};
    \vertex [left=0.5cm of gm] (g2) {};
    \vertex [right=0.5cm of gm] (g20) {};
    \diagram* {
        (a) -- [fermion,thick] (f2) -- [fermion,thick] (c),
        (g2) -- [photon, ultra thick] (f2),
        (f2) -- [photon, ultra thick] (g20)
    };
\end{feynman}
\end{tikzpicture}\kern-0.6em\Big]&=0 .
\end{align}
Imposing the appropriate mass scaling and the desired spin order restricts the ansatz for $\CN_{\rmA,\rm c}$ to a finite number of terms.  Eq.~\eqref{eq:s5cons} then yields a unique solution,
\begin{align}\label{eq:M15}
\mathcal{M}^{(1,5)}_{\rmA,\rm c}&=16\pi \epsbar M  |a|i\Big(\left(-U_1+U_2+U_5-U_6\right)\mathsf{b}_1+
 \left(-U_3+U_4-U_7+U_8\right)\mathsf{b}_2\Big)
\, ,
\end{align}
with the $U_i$ defined below, and their transformation properties under the crossing replacement $p_1\to -p_2$, $p_2\to -p_1$ indicated by the arrows:
\begin{align}
U_1 &= |a|\, a\mdot F_1\mdot v\, a\mdot F_2^*\mdot v\, a\mdot F_1\mdot \widetilde F_2^*\mdot v,
&\longleftrightarrow&&
U_2 &= |a|\, a\mdot F_1\mdot v\, a\mdot F_2^*\mdot v\, a\mdot F_2^*\mdot \widetilde F_1\mdot v,
\nn\\
U_3 &= |a|\, a\mdot F_1\mdot F_2^*\mdot a\, a\mdot F_1\mdot \widetilde F_2^*\mdot v,
&\longleftrightarrow&&
U_4 &= |a|\, a\mdot F_1\mdot F_2^*\mdot a\, a\mdot F_2^*\mdot \widetilde F_1\mdot v,
\nn\\
U_5 &= |a|\, a\mdot F_1\mdot v\, a\mdot F_2^*\mdot v\, a\mdot \widetilde F_1\mdot F_2^*\mdot v,
&\longleftrightarrow&&
U_6 &= |a|\, a\mdot F_1\mdot v\, a\mdot F_2^*\mdot v\, a\mdot \widetilde F_2^*\mdot F_1\mdot v,
\nn\\
U_7 &= |a|^3\, v\mdot F_1\mdot F_2^*\mdot v\, a\mdot F_1\mdot \widetilde F_2^*\mdot v,
&\longleftrightarrow&&
U_8 &= |a|^3\, v\mdot F_1\mdot F_2^*\mdot v\, a\mdot F_2^*\mdot \widetilde F_1\mdot v \, .
\end{align}
Together with the transformation $\epsbar\to -\epsbar$ (since $\omega=p_1\mdot v\to  -p_2\mdot v=-\omega$), it is easy to see that $\mathcal{M}^{(1,5)}_{\rmA,\rm c}$ is crossing symmetric. We note that this contribution is real and, therefore, gives rise to a conservative finite-size effect. The factor of $i$ appears together with the Levi-Civita tensor in $\widetilde F$, so the contribution becomes manifestly real in the coordinate system of Figure~\ref{fig:XYZ}. Moreover, the powers of $|a|$ that appear in the amplitude are always even. They can therefore be written in terms of $a\cdot a$, which arises naturally in the effective field theory amplitude. 

Using eq.~\eqref{eq:Pfg}, one can directly verify that the functions $U_1$ to $U_8$ contain only $l=2$ modes, and hence $\mathcal{M}^{(1,5)}_{\rmA,\rm c}$ only contributes to the $l=2$ modes, in agreement with Table~\ref{tab:fitType1}. Both linear combinations of the $U_i$ associated with $\mathsf{b}_1$ and $\mathsf{b}_2$ in eq.~\eqref{eq:M15} have vanishing helicity-reversing components, so that $\mathcal{M}^{(1,5)}_{\rmA,\rm c}$ is non-vanishing only in the helicity-conserving sector.

\subsection{At Sixth Order in Spin}
At sixth order in spin, black hole perturbation theory yields non-vanishing contributions, starting at $\terms{\epsbar^0}$, in the helicity-conserving sector for $l=2$. By contrast, $\mathcal{M}_{\rmS,\text{fin.}}$ first contributes at $\terms{\epsbar}$. Thus, the difference $f_{\rm diff.}^{(l,m)}\big|_{\epsbar^0 |a|^6}$ is just the result from black hole perturbation theory,
\begin{equation}
\begin{aligned}
f_{\rm diff.}^{(2,-2)}\Big|_{\epsbar^0|a|^6}
&=4 \sqrt{\frac{\pi }{5}} \mathsf{c}_1 \omega ^5 | a| ^6 \sin ^4(\gamma ),~~
f_{\rm diff.}^{(2,-1)}\Big|_{\epsbar^0|a|^6}
=
-8 \sqrt{\frac{\pi }{5}} \mathsf{c}_1 \omega ^5 | a| ^6 \sin ^3(\gamma ) \cos (\gamma ),
\\[6pt]
f_{\rm diff.}^{(2,0)}\Big|_{\epsbar^0|a|^6}
&=
2 \sqrt{\frac{2 \pi }{15}} \omega ^5 | a| ^6 \sin ^2(\gamma ) (3 \mathsf{c}_1 \cos (2 \gamma )+3 \mathsf{c}_1+4 \mathsf{c}_2),
\\[6pt]
f_{\rm diff.}^{(2,1)}\Big|_{\epsbar^0|a|^6}
&=
-2 \sqrt{\frac{\pi }{5}} \omega ^5 | a| ^6 \sin (2 \gamma ) (\mathsf{c}_1 \cos (2 \gamma )+\mathsf{c}_1+4 \mathsf{c}_2),
\\[6pt]
f_{\rm diff.}^{(2,2)}\Big|_{\epsbar^0|a|^6}
&=
\frac{1}{6} \sqrt{\frac{\pi }{5}} \omega ^5 | a| ^6 (3 \mathsf{c}_1 \cos (4 \gamma )+12 \cos (2 \gamma ) (\mathsf{c}_1+4 \mathsf{c}_2)+12 \mathsf{c}_1+44 \mathsf{c}_2+27 \mathsf{c}_3),
\end{aligned}
\end{equation}
where the $\mathsf{c}_i$ are purely imaginary numbers,
\begin{equation}\label{eq:coefsa6e0}
    \begin{aligned}
  \mathsf{c}_1&=\frac{1}{15} i (10 \zeta (2)-92 \zeta (4)+80 \zeta (6)-1), \\
 \mathsf{c}_2&=\frac{1}{120} i (120 \zeta (2)-1152 \zeta (4)+1024 \zeta (6)-1), \\
 \mathsf{c}_3&=\frac{1}{90} i (60 \zeta (2)-600 \zeta (4)+544 \zeta (6)-5)\, .
    \end{aligned}
\end{equation}
Following the same steps as taken at fifth order in spin, we set up the ansatz for $\CN_{\rmA,\rm c}$ and impose the constraint equations, 
\begin{align}
\mathbb{P}_f^{(l,m)}\Big[\kern-0.6em\begin{tikzpicture}[scale=1.5, transform shape, baseline=0.1cm]
\tikzstyle{every node}=[font=\small]
\begin{feynman}
    \vertex (a) {};
    \vertex [right=0.6cm of a] (f2) [myblob2] {$\,\mathbf{0}\,$};
    \vertex [above=0.3cm of f2]{\tiny $6$};
    \vertex [right=0.6cm of f2] (c) {};
    \vertex [above=0.5cm of f2] (gm) {};
    \vertex [left=0.5cm of gm] (g2) {};
    \vertex [right=0.5cm of gm] (g20) {};
    \diagram* {
        (a) -- [fermion,thick] (f2) -- [fermion,thick] (c),
        (g2) -- [photon, ultra thick] (f2),
        (f2) -- [photon, ultra thick] (g20)
    };
\end{feynman}
\end{tikzpicture}\kern-0.6em\Big]&=\begin{cases}{8\pi M} f_{\rm diff.}^{(2,m)} \Big|_{\epsbar^0 |a|^6},& l=2\\
0,&l\neq 2
\end{cases}\, , &\mathbb{P}_g^{(l,m)}\Big[\kern-0.6em\begin{tikzpicture}[scale=1.5, transform shape, baseline=0.1cm]
\tikzstyle{every node}=[font=\small]
\begin{feynman}
    \vertex (a) {};
    \vertex [right=0.6cm of a] (f2) [myblob2] {$\,\mathbf{0}\,$};
    \vertex [above=0.3cm of f2]{\tiny $6$};
    \vertex [right=0.6cm of f2] (c) {};
    \vertex [above=0.5cm of f2] (gm) {};
    \vertex [left=0.5cm of gm] (g2) {};
    \vertex [right=0.5cm of gm] (g20) {};
    \diagram* {
        (a) -- [fermion,thick] (f2) -- [fermion,thick] (c),
        (g2) -- [photon, ultra thick] (f2),
        (f2) -- [photon, ultra thick] (g20)
    };
\end{feynman}
\end{tikzpicture}\kern-0.6em\Big]&=0\, .
\end{align}
The unique solution is an imaginary amplitude which encodes dissipative finite-size effects,
\begin{equation}\label{eq:MII06}
\begin{aligned}
   \mathcal{M}^{(0,6)}_{\rmA,\rm c}={16\pi M }\omega |a|^2 \Big[&
   \mathsf{c}_2 (2 U_{10}+U_{13}+U_{14})+\mathsf{c}_1 (U_{9}-{U_{12}\over 4} )+\mathsf{c}_3 (U_{11}-{U_{12}\over 4} )\Big]\, ,
\end{aligned}
\end{equation}
where the $U_i$ are given by the crossing invariant functions, 
\begin{align}
U_{9} &= \left(a\mdot F_1\mdot F_2^*\mdot a\right)^2,
&
U_{10} &= |a|^2\, a\mdot F_1\mdot F_2^*\mdot v\, a\mdot F_2^*\mdot F_1\mdot v,
\nn\\
U_{11} &= |a|^4\, \left(v\mdot F_1\mdot F_2^*\mdot v\right)^2,&
U_{12} &= |a|^4\, \text{tr}\left(F_1\mdot F_2^*\right)\, v\mdot F_1\mdot F_2^*\mdot v ,
\end{align} 
and two more functions that map to each other under the crossing replacement,
\begin{align}
U_{13} &= |a|^2\, \left(a\mdot F_1\mdot F_2^*\mdot v\right)^2,
&\longleftrightarrow &&
U_{14} &= |a|^2\, \left(a\mdot F_2^*\mdot F_1\mdot v\right)^2.
\end{align}
Combined with the transformation $\omega\rightarrow-\omega$, the amplitude is antisymmetric under the crossing transformation. We have verified that each of these $U_i$ receives contributions exclusively from the $l=2$ modes. Consequently,  $\mathcal{M}_{\rmA,\rm c}^{(0,6)}$ contains only $l=2$ modes. Furthermore, each combination of the $U_i$ enclosed in parentheses in eq.~(\ref{eq:MII06}) yields a vanishing helicity-reversing component.

To complete the low-$\epsbar$ sector at sixth order in spin, we continue to linear order in $\epsbar$ and compute the difference between black hole perturbation theory and $\mathcal{M}_{\rmS,\text{fin.}}$. The difference is nontrivial only for the $l=2,3$ modes. At this order, we must also consider the loop contribution that can be evaluated by gluing $\mathcal{M}^{(0,6)}_{\rmA,\rm c}$ to the spinless tree-level amplitude (see Appendix \ref{app:1loopRes} for the details of this one-loop calculation),
\begin{align}\label{eq:M1Loop}
\mathcal{M}_{\rmA,\text{fin.}}^{1\text{Loop}}\Big|_{\epsbar a^{6}}&=\epsbar\,\big(\pi +i (2 \gamma_{\rm E} -{25\over 12})\big) \mathcal{M}^{(0,6)}_{\rmA,\rm c}+(\text{helicity-reversing part}).
\end{align}
We note that the helicity-reversing part above contains only $l=2$ modes and lines up with $g_{\rm diff.}^{(l,m)}\Big|_{\epsbar |a|^6}$,
\begin{align}\label{eq:spin6g}
\mathbb{P}_g^{(l,m)}\Big[\mathcal{M}_{\rmA,\text{fin.}}^{1\text{Loop}}\Big|_{\epsbar a^{6}}\Big]&\equiv\mathbb{P}_g^{(l,m)}\Big[\kern-0.6em\begin{tikzpicture}[scale=1.5, transform shape, baseline=0.1cm]
\tikzstyle{every node}=[font=\small]
\begin{feynman}
    \vertex (a) {};
    \vertex [right=0.6cm of a] (f2) [myblob2] {$\,\mathbf{0}\,$};
    \vertex [right=0.7cm of f2] (f3) [HV2] {$\,\mathbf{1}\,$};
    \vertex [above=0.3cm of f2]{\tiny $6$};
    \vertex [above=0.3cm of f3]{\tiny $0$};
    \vertex [right=1.3cm of f2] (c) {};
    \vertex [above=0.5cm of f2] (gm) {};
    \vertex [left=0.5cm of gm] (g2) {};
    \vertex [right=1.2cm of gm] (g20) {};
    \vertex [right=0.35cm of f2] (cut1);
    \vertex [below=0.15cm of cut1] (cut1u);
    \vertex [above=0.1cm of cut1] (cut1b);
    \vertex [above=0.4cm of cut1b] (cut1bb);
    \diagram* {
        (a) -- [fermion,thick] (f2) -- [fermion,thick] (f3)-- [fermion,thick] (c),
        (f2)-- [photon,ultra thick,out=35,in=145,looseness=0.5,min distance=0.2cm] (f3),
        (g2) -- [photon, ultra thick] (f2),
        (f3) -- [photon, ultra thick] (g20),
         (cut1u)--[red,thick] (cut1b),
      (cut1b)--[dashed,red,thick] (cut1bb)
    };
\end{feynman}
\end{tikzpicture}\kern-0.6em\Big]+\mathbb{P}_g^{(l,m)}\Big[\kern-0.6em\begin{tikzpicture}[scale=1.5, transform shape, baseline=0.1cm]
\tikzstyle{every node}=[font=\small]
\begin{feynman}
    \vertex (a) {};
    \vertex [right=0.6cm of a] (f2) [HV2] {$\,\mathbf{1}\,$};
    \vertex [right=0.7cm of f2] (f3) [myblob2] {$\,\mathbf{0}\,$};
    \vertex [above=0.3cm of f2]{\tiny $0$};
    \vertex [above=0.3cm of f3]{\tiny $6$};
    \vertex [right=1.3cm of f2] (c) {};
    \vertex [above=0.5cm of f2] (gm) {};
    \vertex [left=0.5cm of gm] (g2) {};
    \vertex [right=1.2cm of gm] (g20) {};
    \vertex [right=0.35cm of f2] (cut1);
    \vertex [below=0.15cm of cut1] (cut1u);
    \vertex [above=0.1cm of cut1] (cut1b);
    \vertex [above=0.4cm of cut1b] (cut1bb);
    \diagram* {
        (a) -- [fermion,thick] (f2) -- [fermion,thick] (f3)-- [fermion,thick] (c),
        (f2)-- [photon,ultra thick,out=35,in=145,looseness=0.5,min distance=0.2cm] (f3),
        (g2) -- [photon, ultra thick] (f2),
        (f3) -- [photon, ultra thick] (g20),
         (cut1u)--[red,thick] (cut1b),
      (cut1b)--[dashed,red,thick] (cut1bb)
    };
\end{feynman}
\end{tikzpicture}\kern-0.6em\Big]\nn\\
&=8\pi M g_{\rm diff.}^{(l,m)}\Big|_{\epsbar |a|^6}\, . 
\end{align}
Thus, the constraints on the contact contribution at sixth order in spin and linear order in $\epsbar$,
\begin{align}\label{eq:spin6}
\mathbb{P}_f^{(l,m)}\Big[\kern-0.6em\begin{tikzpicture}[scale=1.5, transform shape, baseline=0.1cm]
\tikzstyle{every node}=[font=\small]
\begin{feynman}
    \vertex (a) {};
    \vertex [right=0.6cm of a] (f2) [myblob2] {$\,\mathbf{1}\,$};
    \vertex [above=0.3cm of f2]{\tiny $6$};
    \vertex [right=0.6cm of f2] (c) {};
    \vertex [above=0.5cm of f2] (gm) {};
    \vertex [left=0.5cm of gm] (g2) {};
    \vertex [right=0.5cm of gm] (g20) {};
    \diagram* {
        (a) -- [fermion,thick] (f2) -- [fermion,thick] (c),
        (g2) -- [photon, ultra thick] (f2),
        (f2) -- [photon, ultra thick] (g20)
    };
\end{feynman}
\end{tikzpicture}\kern-0.6em\Big]&=\begin{cases}\vspace{0.4em}8\pi M f_{\rm diff.}^{(l,m)}\Big|_{\epsbar |a|^6} -\epsbar\big(\pi+i(2 \gamma_{\rm E} -{25\over 12})\big)\, \mathbb{P}_f^{(l,m)}\Big[\kern-0.6em\begin{tikzpicture}[scale=1.5, transform shape, baseline=0.1cm]
\tikzstyle{every node}=[font=\small]
\begin{feynman}
    \vertex (a) {};
    \vertex [right=0.6cm of a] (f2) [myblob2] {$\,\mathbf{0}\,$};
    \vertex [above=0.3cm of f2]{\tiny $6$};
    \vertex [right=0.6cm of f2] (c) {};
    \vertex [above=0.5cm of f2] (gm) {};
    \vertex [left=0.5cm of gm] (g2) {};
    \vertex [right=0.5cm of gm] (g20) {};
    \diagram* {
        (a) -- [fermion,thick] (f2) -- [fermion,thick] (c),
        (g2) -- [photon, ultra thick] (f2),
        (f2) -- [photon, ultra thick] (g20)
    };
\end{feynman}
\end{tikzpicture}\kern-0.6em\Big]\,,\,& l=2,3\\
0\,,&l\geq 4\\[0.3em]
\end{cases}\nn\\
\mathbb{P}_g^{(l,m)}\Big[\kern-0.6em\begin{tikzpicture}[scale=1.5, transform shape, baseline=0.1cm]
\tikzstyle{every node}=[font=\small]
\begin{feynman}
    \vertex (a) {};
    \vertex [right=0.6cm of a] (f2) [myblob2] {$\,\mathbf{1}\,$};
    \vertex [above=0.3cm of f2]{\tiny $6$};
    \vertex [right=0.6cm of f2] (c) {};
    \vertex [above=0.5cm of f2] (gm) {};
    \vertex [left=0.5cm of gm] (g2) {};
    \vertex [right=0.5cm of gm] (g20) {};
    \diagram* {
        (a) -- [fermion,thick] (f2) -- [fermion,thick] (c),
        (g2) -- [photon, ultra thick] (f2),
        (f2) -- [photon, ultra thick] (g20)
    };
\end{feynman}
\end{tikzpicture}\kern-0.6em\Big]&=0 \, .
\end{align}
produce a unique solution in terms of five independent coefficients
\begin{equation}
    \begin{aligned}
\mathsf d_1&=-\frac{8}{135} (64 \zeta (3)-64 \zeta (5)-7), \\
\mathsf d_2&=-\frac{2}{135} (4 \zeta (3)-16 \zeta (5)+5), \\
\mathsf d_3&=\frac{107\zeta(2)}{630}
+\frac{56\zeta(3)}{15}
+\frac{428\zeta(6)}{315}
+\frac{56\zeta(7)}{3}
-\frac{202\zeta(5)}{9}
-\frac{2461\zeta(4)}{1575}
-\frac{183}{700},\\
\mathsf d_4&=\frac{428\zeta(2)}{105}
+\frac{10976\zeta(3)}{135}
+\frac{54784\zeta(6)}{1575}
+\frac{7168\zeta(7)}{15}
-\frac{75392\zeta(5)}{135}
-\frac{6848\zeta(4)}{175}
-\frac{25661}{9450},\\
\mathsf d_5&=-\frac{107\zeta(2)}{45}
+\frac{36166\zeta(4)}{1575}
+\frac{44564\zeta(5)}{135}
-\frac{12656\zeta(7)}{45}
-\frac{6656\zeta(3)}{135}
-\frac{96728\zeta(6)}{4725}
+\frac{8404}{4725},\\
    \end{aligned}
\end{equation}
so that
\begin{align}\label{eq:M16}
\mathcal{M}^{(1,6)}_{\rmA,\rm c}&=16\pi \epsbar  M |a|\Big[\mathsf d_3 \big(\frac{1}{2} U_{12} \omega  | a| -U_{23} a\cdot p_2-U_{24} a\cdot p_1+U_{19}+U_{20}\big)\nn\\
&+\mathsf d_1 \big(-U_{22} \omega  | a| +\frac{1}{2} U_{12} \omega  | a| -U_{23} a\mdot p_2-U_{24} a\mdot p_1+U_{15}+U_{16}\big)\nn\\
&+\mathsf d_2 \big(U_{22} \omega  | a| -\frac{1}{2} U_{12} \omega  | a| +U_{23} a\mdot p_2+U_{24} a\mdot p_1+U_{17}+U_{18}\big)\nn\\
&+\frac{1}{4} \mathsf d_4 \big(4 U_{21}-U_{22}\big) \omega  | a| +\frac{1}{4} \mathsf d_5 \big(4 U_{11}-U_{12}\big) \omega  |a|\Big]\, .
\end{align}
This amplitude is real and therefore describes conservative finite-size effects. The crossing invariant functions in eq.~\eqref{eq:M16} are given by
\begin{align}
U_{21} &= |a|^2\, v\mdot F_1\mdot F_2^*\mdot v\, a\mdot F_1\mdot v\, a\mdot F_2^*\mdot v,
& U_{22} &= |a|^2\, \text{tr}\left(F_1\mdot F_2^*\right)\, a\mdot F_1\mdot v\, a\mdot F_2^*\mdot v,
\end{align}
while the other $U$ functions transform under the crossing replacement as 
\begin{align}
U_{15} &= |a|^3\, a\mdot F_2^*\mdot v\, a\mdot F_1\mdot F_2^*\mdot v\, p_2\mdot F_1\mdot v,
&\overset{-}{\longleftrightarrow}&&
U_{16} &= |a|^3\, a\mdot F_1\mdot v\, a\mdot F_2^*\mdot F_1\mdot v\, p_1\mdot F_2^*\mdot v,
\nn\\
U_{17} &= |a|^3\, a\mdot F_1\mdot v\, a\mdot F_1\mdot F_2^*\mdot v\, p_1\mdot F_2^*\mdot v,
&\overset{-}{\longleftrightarrow}&&
U_{18} &= |a|^3\, a\mdot F_2^*\mdot v\, a\mdot F_2^*\mdot F_1\mdot v\, p_2\mdot F_1\mdot v, \nn\\
U_{19} &= |a|\, a\mdot F_1\mdot F_2^*\mdot a\, a\mdot F_1\mdot p_2\, a\mdot F_2^*\mdot v,
&\overset{-}{\longleftrightarrow}&&
U_{20} &= |a|\, a\mdot F_1\mdot F_2^*\mdot a\, a\mdot F_2^*\mdot p_1\, a\mdot F_1\mdot v,
\nn\\
U_{23} &= |a|\, a\mdot F_1\mdot F_2^*\mdot a\, a\mdot F_1\mdot F_2^*\mdot v,
& \longleftrightarrow&&
U_{24} &= |a|\, a\mdot F_1\mdot F_2^*\mdot a\, a\mdot F_2^*\mdot F_1\mdot v\,,
\end{align}
where $U_i \overset{-}{\longleftrightarrow} U_j$ denotes the transformation
$U_i\rightarrow -U_j,\, U_j\rightarrow -U_i$.

It is easy to see that $\mathcal{M}^{(1,6)}_{\rmA,\rm c}$ is crossing symmetric. Furthermore, $U_{11}$, $U_{12}$, $U_{16}$, $U_{17}$, $U_{20}$, $U_{21}$, $U_{22}$, and $U_{24}\,a\mdot p_1$ contain only $l=2$ modes, whereas the remaining terms in eq.~\eqref{eq:M16} contain both $l=2$ and $l=3$ modes. Hence $\mathcal{M}^{(1,6)}_{\rmA,\rm c}$ can contribute only to $l=2$ and $l=3$ modes, as expected from Table~\ref{tab:fitType1}. Each combination of the $U_i$ in parentheses in eq.~(\ref{eq:M16}) yields a vanishing helicity-reversing component, confirming that $\mathcal{M}^{(1,6)}_{\rmA,\rm c}$ is non-vanishing only in the helicity-conserving sector.

We have thus uniquely determined the tree-level Compton amplitude through sixth order in spin and to linear order in $\epsbar$,
\begin{align}\label{eq:all-loop Compton2}
        \mathcal{M}^{\rm tree}_{\rm Kerr}&= 32\pi G M^2\Big(\frac{\mathcal{N}_0\mathcal{N}_a}{4\omega^2x^{2}}-{1\over 4M^2\omega^2}\mathcal{N}_r+\CN_{\rmS,\rm c}\Big)+ \mathcal{M}^{(1,5)}_{\rmA,\rm c}+\mathcal{M}^{(0,6)}_{\rmA,\rm c}+\mathcal{M}^{(1,6)}_{\rmA,\rm c}+\mathcal{O}(\epsbar^2,|a|^7)\,,
    \end{align}
where $\mathcal{O}(\epsbar^2,|a|^7)$ denotes terms of order $\mathcal{O}(\epsbar^2)$ or $\mathcal{O}(|a|^7)$ and higher.
The appearance of alternating transcendental functions in the $\mathcal{M}_{\rmA}$ terms reflects the nontrivial analytic structure inherited from the Mano-Suzuki-Takasugi solution. We emphasize that, apart from the double-copy contribution, which has the exponential structure for the helicity-reversing amplitude, all other contributions to the tree-level amplitude are helicity-conserving. This provides a direct verification, through sixth order in spin, of the exponential structure suggested for the helicity-reversing amplitude~\cite{Arkani-Hamed:2017jhn,Guevara:2018wpp,Bautista:2021wfy,Aoude:2020onz,Bautista:2023sdf,Bautista:2024agp}. 

\subsection{Absorption Effects from the Leading Contact Term and its Tail at Sixth Order in Spin }
\label{sec:sextic-absorption}
\begin{figure}[h]
\centering
\begin{tikzpicture}[
    >=Stealth,
    line cap=round,
    line join=round,
    scale=0.8
]

\coordinate (O) at (0,0);

\draw[->, very thick] (O) -- (0,4.2)
    node[above] {$Z_B$};

\draw[->, very thick] (O) -- (3.7,0)
    node[right] {$Y$};

\draw[->, very thick] (O) -- (-2.0,-1.9)
    node[below left] {$X_B$};
\draw[dashed, very thick] (O) -- (2.0,1.9)
    ;
\draw[->, very thick, brown!80!black]
    (O) -- (0,3.25)
    node[right=0.12cm] {$\vec a$};

\coordinate (A) at (-0.85,2.45);

\draw[->, very thick, blue]
    (O) -- (A)
    node[above left] {$\vec p_1$};
\draw[dashed, very thick, blue]
    (A) -- (-0.85,-0.85);

\coordinate (P) at (2.15,1.15);

\draw[->, very thick, red]
    (O) -- (P)
    node[above right] {$\vec p_2$};

\coordinate (Q) at (2.15,-1.05);

\draw[dashed, red, thick]
    (P) -- (Q);

\draw[dashed, red, thick]
    (O) -- (Q);

\draw[
    blue!80!black,
    thick,
    -,
    domain=0:0
] (0,2.15) arc[start angle=90,end angle=122,radius=1.25];

\node[blue!80!black] at (-0.42,2.35) {$\gamma$};

\draw[
    red,
    thick,
    -,
    domain=0:0
] (0,1.45) arc[start angle=90,end angle=28,radius=1.45];

\node[red] at (0.68,1.07) {$\theta_B$};

\draw[
    red,
    thick,
    -,
    domain=0:0
] (-0.65,-0.65)
    arc[start angle=225,end angle=327,radius=1.05];

\node[red] at (0.05,-1.18) {$\phi_B$};

\end{tikzpicture}
\caption{Kinematic configuration in a coordinate frame where the spin is aligned with $Z_B$.}
\label{fig:BL}
\end{figure}
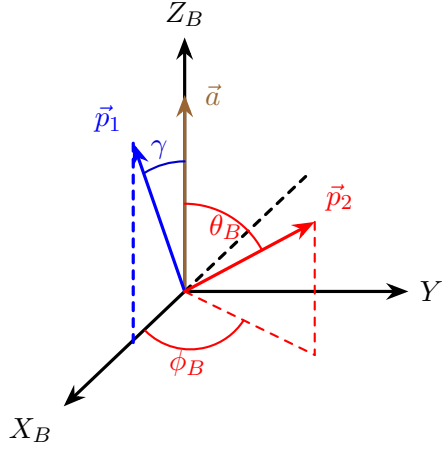

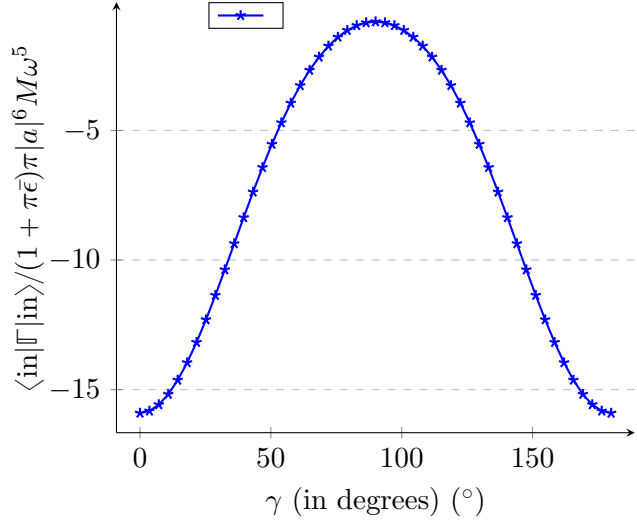
\begin{figure}
\centering
\begin{tikzpicture}[scale=1.0]
\begin{axis}[
title={},   axis lines=left,
			xlabel=$\gamma\ \text{(in  degrees)} \ (^\circ)$, 
			ylabel=$\langle {\rm in}|\bG|{\rm in}\rangle/ (1+\pi \epsbar)\pi |a|^6 M\omega^5$, ylabel style={rotate=0,at={(-0.12,0.5)},anchor=south},
			legend style={at={(0.25,1.)},anchor=north},
			ymajorgrids=true,
			grid style=dashed,
            every axis plot/.append style={thick},enlargelimits=0.05]
\addplot[color=blue, mark=star] coordinates {(0,-15.9075) (3.6,-15.8251) (7.2,-15.58) (10.8,-15.1794) (14.4,-14.6347) (18,-13.961) (21.6,-13.1768) (25.2,-12.303) (28.8,-11.3618) (32.4,-10.3763) (36,-9.36932) (39.6,-8.36266) (43.2,-7.37633) (46.8,-6.42801) (50.4,-5.53268) (54,-4.70229) (57.6,-3.94581) (61.2,-3.26927) (64.8,-2.67606) (68.4,-2.16736) (72,-1.74259) (75.6,-1.4) (79.2,-1.13719) (82.8,-0.951682) (86.4,-0.841334) (90,-0.804718) (93.6,-0.841334) (97.2,-0.951682) (100.8,-1.13719) (104.4,-1.4) (108,-1.74259) (111.6,-2.16736) (115.2,-2.67606) (118.8,-3.26927) (122.4,-3.94581) (126,-4.70229) (129.6,-5.53268) (133.2,-6.42801) (136.8,-7.37633) (140.4,-8.36266) (144,-9.36932) (147.6,-10.3763) (151.2,-11.3618) (154.8,-12.303) (158.4,-13.1768) (162,-13.961) (165.6,-14.6347) (169.2,-15.1794) (172.8,-15.58) (176.4,-15.8251) (180,-15.9075)};
\addlegendentry{}
\end{axis}
\end{tikzpicture}
\caption{The absorption coefficient at sixth order in spin as a function of the angle $\gamma$ between the incoming wave and the spin direction. }
\label{fig:abs}
\end{figure}

We now investigate the physical character of the leading sixth-order-in-spin contact term identified above. As we show below, this term contributes to the dissipative effects encoded in the non-unitary $\bS$-matrix of the effective field theory. The loss of unitarity can be characterized by the absorption operator, 
\begin{equation}\label{eq:abs-op}
    \bG\equiv \mathbb{1} - \bS^{\dagger}\bS= i\,\bT^\dagger - i\,\bT - \bT^\dagger\bT,
\end{equation}
where the $\bT$-matrix is defined by $\bS = \mathbb{1} + i\bT$. We work in a coordinate system $\{X_B,Y,Z_B\}$ with the $Z_B$-axis aligned with the spin vector $\Vec a$. Without loss of generality, $\Vec{p}_1$ lies in the $X_B$-$Z_B$ plane and forms an angle $\gamma$ with $\Vec{a}$ while $\Vec{p}_2$ is parametrized by the polar and azimuthal angles $(\theta_B,\phi_B)$, as shown in Figure~\ref{fig:BL}. 
The incoming state is normalized and given by,
\begin{equation}
|{\rm in}\rangle=|\gamma,0,+\rangle. 
\end{equation}
The absorption coefficient is then given by, 
\begin{align}
    \langle {\rm in}|\bG|{\rm in}\rangle &=  2\rm Im  \Big(\langle {\rm in}|\bT|{\rm in}\rangle\Big)- \sum_{\rm out}  \langle {\rm in}| \bT^\dagger  | {\rm out}\rangle \langle {\rm out}| \bT |{\rm in}\rangle \, \nn\\
    &=2{\rm Im} \Big( \langle {\rm in}|\bT|{\rm in}\rangle\Big)+\terms{ \epsbar^2,|a|^7} \, ,
\end{align}
where $
|{\rm out}\rangle =|\theta_B,\phi_B,\pm\rangle\, $. The second term contributes only at higher orders and can therefore be neglected. Since the forward matrix element \(\langle {\rm in}|\bT|{\rm in}\rangle\) is obtained by taking the outgoing direction to coincide with the incoming direction, we take
\begin{align}
{\rm Im } \Big(\langle {\rm in}|\bT|{\rm in}\rangle\Big)&=\rm Im  \Big(\langle \theta_B,\phi_B,+|\bT|\gamma,0,+\rangle\Big)\Big|_{\theta_B\to \gamma}^{\phi_B\to 0}\equiv {\rm Im } \Big(\mathcal{M}_{\text{fin.}}(2e^{-\gamma_{\rm E}}\omega, \varepsilon_1^+, \varepsilon_2^{+*})\Big)\Big|_{\theta_B\to \gamma}^{\phi_B\to 0}\, .
\end{align}
Strictly speaking, in gravitational scattering the forward matrix element is not well-defined because of the forward divergence of the  gravitational Compton amplitude. However, at the order considered in this work (i.e. $\mathcal O(\bar\epsilon^1)$), all forward-divergent contributions are purely real. Consequently, at this order, the imaginary part of the forward matrix element remains well-defined.
We stress that the forward matrix element should be understood in terms of the finite part of the scattering amplitude in the effective field theory, \(\mathcal{M}_{\mathrm{fin.}}\). In particular, the universal Weinberg infrared phase is an overall phase that leaves \(|\mathcal{M}|^2\) invariant but changes the imaginary part of the forward amplitude. Therefore, to establish the correspondence between the effective-field-theory amplitude and the quantum-mechanical scattering \(\bT\)-matrix, the unphysical Weinberg phase must be removed first. The absorption coefficient is therefore given by
\begin{align}\label{eq:absAmp}
    \langle {\rm in}|\bG|{\rm in}\rangle &= 2{\rm Im }\Big(\mathcal{M}^{(0,6)}_{\rmA,\rm c}( \varepsilon^+_1, \varepsilon_2^{+*})+ \big(\mathcal{M}_{\rmA,\text{fin.}}^{1\text{Loop}}(2e^{-\gamma_{\rm E}}\omega, \varepsilon^+_1, \varepsilon_2^{+*})\Big|_{\epsbar a^{6}}\big)\Big)\Big|_{\theta_B\to \gamma}^{\phi_B\to 0}+\terms{ \epsbar^2,|a|^7}) \, \nn\\
    &=2{\rm Im }\Big((1+\pi\epsbar)\mathcal{M}^{(0,6)}_{\rmA,\rm c}( \varepsilon^+_1, \varepsilon_2^{+*})\Big)\Big|_{\theta_B\to \gamma}^{\phi_B\to 0}+\terms{ \epsbar^2,|a|^7})\nn\\
    &=(1+\pi \epsbar)\pi |a|^6 M\omega^5
\Big(
16 i \mathsf c_1 \sin ^2(\gamma ) (\cos (2 \gamma )+3)-{64\over 3} i \mathsf c_2 (3 \cos (2 \gamma )+5)-{32\over 9}
\Big)\nn\\
&+\terms{ \epsbar^2,|a|^7})\,.
\end{align}
The absorptive contribution is dressed by the universal factor $1+\pi\bar\epsilon$, with the term proportional to $\pi\bar\epsilon$ reproduced by the one-loop tail contribution. The absolute value of the absorption coefficient peaks when the incoming wave is aligned or anti-aligned with the spin direction and is minimal when they are orthogonal, as shown in Figure~\ref{fig:abs}.

The absorption effects, in classical gravity, can also be obtained from the cross-section formula~\cite{Dolan:2008kf},  
\begin{equation}\label{eq:absCS}
   \sigma_{\rm abs} \equiv \frac{4\pi^2}{\omega^2}\sum_{l=2}^{\infty}\sum_{m=-l}^{l}|{}_{-2}S_{lm}(\gamma,0;|a|\omega)|^2\mathbb{\Gamma}_{lm},
\end{equation}
where the transmission factor $\mathbb{\Gamma}_{l m}
    \equiv
    1-\left|e^{2 i \delta_{l m}^{\pm}}\right|^2 $ is related to the imaginary part of the phase shift $\delta_{l m}^{\pm}$. (Note that the norm $\left|e^{2 i \delta_{l m}^{\pm}}\right|$ does not depend on the helicity configuration~\cite{Dolan:2008kf}).
Expanding the phase shift from the Mano-Suzuki-Takasugi solution up to sixth order in spin, we find that it is nontrivial only for
$l=2$.
\begin{equation}\label{eq:SNorm}
    \mathbb{\Gamma}_{2m}=- \rho_m\,|a|^6\omega^6
    \left(
        1+\pi\bar\epsilon
    \right)
    +
    \terms{\epsbar^2, |a|^7},
\end{equation}
where
\begin{align}\label{eq:rhos}
    \rho_{\pm2}&=
   \frac{4}{45}+\frac{64 i \mathsf c_2}{15},&
    \rho_{\pm1}
    &=
    -\frac{8i \mathsf c_1}{5}  +\frac{16 i \mathsf c_2}{15}+\frac{4}{45},&
    \rho_{0}
    &=
    \frac{4}{45}\,.
\end{align}
Substituting \eqref{eq:rhos} into \eqref{eq:SNorm} and comparing eqs.~\eqref{eq:absAmp} and \eqref{eq:absCS}, we verify that the absorption coefficient defined in effective field theory is consistent with the absorption cross section obtained from classical gravitational scattering, at least through $\terms{\epsbar^2, |a|^7}$,
\begin{align}
\langle {\rm in}|\bG|{\rm in}\rangle = { 8 M \omega} \sigma_{\rm abs}\, .
\end{align}
The overall normalization factor arises from different conventions used to define the two quantities. This agreement provides an additional nontrivial consistency check: the leading sixth order in spin contact term obtained in eq.~(\ref{eq:MII06}) agrees with the results reported in refs.~\cite{Saketh:2023bul,Bautista:2024emt} (see also absorption effects from the scalar Compton amplitude in \cite{Jones:2023ugm}).

\section{Conclusion and Outlook}\label{sec:CO}
In this work, we have developed a new framework for computing the gravitational Kerr-Compton amplitude within a post-Minkowskian effective field theory, combining a previously developed method based on consistent physical factorization properties with new input from black hole perturbation theory. Concretely, we obtain the tree-level Compton amplitude through sixth order in spin and to linear order in $\epsbar$. With new data now available, these effective tree amplitudes could yield more accurate spin-dependent contributions to the bending angle and other observables.

As expected, at sufficiently high spin order, terms with non-positive powers of $\epsbar$ must be included in the effective Kerr-Compton amplitude. These new types of terms give rise to loop contributions at higher orders in $\epsbar$ and are therefore also relevant for determining the tree-level contact terms at those orders. For example, to determine the contact terms at seventh order in spin, we need to compute two-loop amplitudes such as 
\begin{align}
\begin{tikzpicture}[scale=1.5, transform shape, baseline=0.1cm]
\tikzstyle{every node}=[font=\small]
\begin{feynman}
    \vertex (a) {};
    \vertex [right=0.6cm of a] (f2) [HV2] {$\,\mathbf{1}\,$};
    \vertex [right=0.7cm of f2] (f3) [myblob2] {-$\mathbf{1}$};
    \vertex [right=0.7cm of f3] (f4) [HV2] {$\,\mathbf{1}\,$};
    \vertex [above=0.3cm of f2]{\tiny $0$};
    \vertex [above=0.3cm of f4]{\tiny $0$};
    \vertex [above=0.3cm of f3]{\tiny $7$};
    \vertex [right=0.6cm of f4] (c) {};
    \vertex [above=0.5cm of f2] (gm) {};
    \vertex [left=0.5cm of gm] (g2) {};
    \vertex [right=1.8cm of gm] (g20) {};
    \vertex [right=0.35cm of f2] (cut1);
    \vertex [below=0.15cm of cut1] (cut1u);
    \vertex [above=0.1cm of cut1] (cut1b);
    \vertex [above=0.4cm of cut1b] (cut1bb);
    \vertex [right=0.35cm of f3] (cut11);
    \vertex [below=0.15cm of cut11] (cut11u);
    \vertex [above=0.1cm of cut11] (cut11b);
    \vertex [above=0.4cm of cut11b] (cut11bb);
    \diagram* {
        (a) -- [fermion,thick] (f2) -- [fermion,thick] (f3)-- [fermion,thick] (f4)-- [fermion,thick] (c),
        (f2)-- [photon,ultra thick,out=35,in=145,looseness=0.5,min distance=0.2cm] (f3),(f3)-- [photon,ultra thick,out=35,in=145,looseness=0.5,min distance=0.2cm] (f4),
        (g2) -- [photon, ultra thick] (f2),
        (f4) -- [photon, ultra thick] (g20),
         (cut1u)--[red,thick] (cut1b),
      (cut1b)--[dashed,red,thick] (cut1bb), (cut11u)--[red,thick] (cut11b),
      (cut11b)--[dashed,red,thick] (cut11bb)
    };
\end{feynman}
\end{tikzpicture}.
\end{align}

More broadly, extending this framework to higher orders in spin and $\bar\epsilon$ could provide access to new finite-size effects, including tidal contributions. As shown in Table~\ref{tab:oderingContact}, the tidal contribution at fifth post-Minkowskian order is verified to vanish at zeroth~\cite{Brunello:2026lzf} and quartic orders in spin. Whether this also holds at linear, quadratic, and cubic orders in spin is an open question that can be settled by computing the spinning Compton amplitude at fifth post-Minkowskian order and matching it to black-hole perturbation theory. Another promising direction is to investigate finite-size effects in the nonlinear response of black holes~\cite{Lagos:2022otp,Mitman:2022qdl,Kehagias:2023ctr,Cardoso:2026llh,Bourg:2024jme,Bucciotti:2026jhr}. This requires generalized effective Compton amplitudes with three or more external graviton legs
\cite{Brandhuber:2021kpo,Brandhuber:2021bsf,Bjerrum-Bohr:2023jau,Das:2026cae}.

Finally, it will be important to develop a deeper physical understanding of the effective Compton amplitude. One possible approach is to explore whether a fundamental action for an extended body could reproduce the transcendental structures of the effective amplitude directly; see, for example, \cite{Bjerrum-Bohr:2025lpw}. We leave this and other related questions for future work.

\section*{Acknowledgements}
We thank Dogan Akpinar, Simon Caron-Huot, Paolo Di Vecchia, and Carl Jordan Eriksen for helpful discussions. The work of N.E.J.B.-B., G.C., K.P., and N.S. was supported in part by the Center of Gravity, a Center of Excellence funded by the Danish National Research Foundation under grant No.~184, and in part by the VILLUM Foundation (grant no.~VIL37766) as well as the European Union's Horizon 2020 research and innovation program under the Marie Sklodowska-Curie grant agreement Nos.~101299389 and 101131233. G.C. also acknowledges funding from the European Union's Horizon 2020 research and innovation program under the Marie Sklodowska-Curie grant agreement No.~847523, INTERACTIONS. C.S. was supported by the Guangzhou Elite Project through grant number JY202211. K.P. acknowledges funding from the Deutsche Forschungsgemeinschaft (DFG, German Research Foundation) - Projektnummer 417533893/GRK2575 {\it Rethinking Quantum Field Theory}. This work is also supported in part by the \href{https://www.sfi.org.bm/}{Simons Foundation International} and the \href{https://www.simonsfoundation.org/}{Simons Foundation} through Simons Foundation grant SFI-MPS-BH-00012593-11.

\appendix
\section{Coordinate Transformations in Black Hole Perturbation Theory}\label{app:BHPT}
The wave-scattering process in black hole perturbation theory is naturally described in the black hole's rest frame, where $v^\mu=(1,0,0,0)$. In this frame, the results of black hole perturbation theory are conventionally expressed in the coordinate system $(X_BYZ_B)$ shown in Figure~\ref{fig:BL}. However, for comparison with the amplitude ansatz introduced in Section~\ref{sec:contactTerms}, it is more convenient to work in the $(XYZ)$-coordinates of Figure~\ref{fig:XYZ}. The two coordinate systems are related by a rotation about the $Y$-axis by an angle $\gamma$ (see also the related coordinate rotations considered in refs.~\cite{Glampedakis:2001cx,Bautista:2021wfy}). To carry out the comparison, we therefore need to express the black hole perturbation theory results in the rotated coordinate system. In this appendix, we explain how this transformation is performed.

The scattering functions $f_{B}(\theta_B,\phi_B)$ and $g_{B}(\theta_B,\phi_B)$, derived from black hole perturbation theory, are expressed in terms of the spin-weighted spheroidal harmonic functions,
\begin{align}
f_B(\theta_B,\phi_B)
&=
\frac{\pi e^{-i\Phi}}{i\omega}\sum_{l=2}^{\infty}
\sum_{m=-l}^{l}
{}_{-2}S_{lm}(\theta_B,\phi_B;|a|\omega)\,
{}_{-2}S_{lm}(\gamma,0;|a|\omega)\Delta^{+}_{lm},
\\[1em]
g_B(\theta_B,\phi_B)
&=\frac{\pi e^{-i\Phi}}{i\omega}
\sum_{l=2}^{\infty}
\sum_{m=-l}^{l}
{}_{-2}S_{lm}(\pi-\theta_B,\phi_B;|a|\omega)\, {}_{-2}S_{lm}(\gamma,0;|a|\omega)\, (-1)^{l+m} \Delta^{-}_{lm}\, .
\end{align}

To express the amplitudes $f_B(\theta_B,\phi_B)$ and $g_B(\theta_B,\phi_B)$ in the $(XYZ)$-coordinates, we begin by expanding ${}_{-2}S_{lm}$ in terms of spin-weighted spherical harmonics,
\begin{align}
\label{eq:expandf}
{}_{-2}S_{lm}(\theta_B,\phi_B;|a|\omega)
=
\sum_{l'=2}^{\infty}
C_{l'}^{(l,m)}(|a|\omega)\,
{}_{-2}Y_{l'm}(\theta_B,\phi_B),
\end{align}
where the coefficients $C_{l'}^{(l,m)}$ are determined by the overlap between the spheroidal and spherical harmonics,
\begin{align}
\label{eq:overlapCoeff}
C_{l'}^{(l,m)}(|a|\omega)
=
\int \sin\theta_B\,d\theta_B\,d\phi_B\,
{}_{-2}Y_{l'm}^{*}(\theta_B,\phi_B)\,
{}_{-2}S_{lm}(\theta_B,\phi_B;|a|\omega).
\end{align}

The spin-weighted spherical harmonics transform under the rotation from the $(X_B Y Z_B)$-coordinates to the $(X Y Z)$-coordinates as follows
\begin{align}
\label{eq:rotatedSpherical}
{}_{-2}Y_{l'm}(\theta_B,\phi_B)
=
\sum_{m'=-l'}^{l'}
D^{l'}_{m'm}(\gamma)\,
{}_{-2}Y_{l'm'}(\theta,\phi),
\end{align}
where $D^{l'}_{m'm}(\gamma)
\equiv
D^{l'}_{m'm}(0,\gamma,0)\,$
is the Wigner $D$-matrix~\cite{Walet:2026ydt} associated with a rotation about the $Y$-axis\footnote{We use \textsc{Mathematica}'s convention for the Wigner \(D\)-function.}.

Substituting \eqref{eq:rotatedSpherical} into \eqref{eq:expandf}, we obtain the rotated spheroidal harmonic quoted in the main text,
\begin{align}\label{eq:Srotf}
{}_{-2}\widehat{S}_{lm}(\theta,\phi;|a|\omega)
\equiv
\sum_{l'=2}^{\infty}\sum_{m'=-l'}^{l'}
C_{l'}^{(l,m)}(|a|\omega)\,
D^{l'}_{m'm}(\gamma)\,
{}_{-2}Y_{l'm'}(\theta,\phi).
\end{align}

The same procedure can be applied to the helicity-reversing $g$. In that case, we expand the spheroidal harmonics ${}_{-2}S_{lm}(\pi-\theta_B,\phi_B;|a|\omega)$ in a basis of spin-weighted spherical harmonics ${}_{-2}Y_{lm}(\pi-\theta_B,\phi_B)$. To rotate these reflected spherical harmonics, we use the reflection identity,
\begin{equation}
    {}_{-2}Y_{lm}(\pi-\theta_B,\phi_B) = (-1)^{l+m}{}_{+2}Y_{lm}(\theta_B,\phi_B),
\end{equation}
and transform ${}_{+2}Y_{lm}(\theta_B,\phi_B)$ as before,
\begin{equation}
\begin{aligned}
    {}_{-2}Y_{lm}(\pi-\theta_B,\phi) &= (-1)^{l+m}\sum_{m'}D^{l}_{m'\,m}(\gamma)\,{}_{2}Y_{lm'}(\theta,\phi)\\
    &= \sum_{m'}(-1)^{m'+m}D^{l}_{m'\,m}(\gamma)\,{}_{-2}Y_{lm'}(\pi-\theta,\phi).
\end{aligned}
\end{equation}
One then obtains
\begin{equation}
    {}_{-2}\widecheck{S}_{lm}(\pi-\theta,\phi;|a|\omega)
\equiv
\sum_{l'=2}^{\infty}\sum_{m'=-l'}^{l'}
(-1)^{m'+m}C_{l'}^{(l,m)}(|a|\omega)\,
D^{l'}_{m'\,m}(\gamma)\,{}_{-2}Y_{l'm'}(\pi-\theta,\phi).
\end{equation}

We can now rewrite equations \eqref{eq:fBL} and \eqref{eq:gBL} as
\begin{align}
f_{\rm B}(\theta,\phi)
&=
\sum_{l'=2}^{\infty}
\sum_{m'=-l'}^{l'}
f^{(l'm')}_{\rm B}\,
{}_{-2}Y_{l'm'}(\theta,\phi),
\\
g_{\rm B}(\theta,\phi)
&=
\sum_{l'=2}^{\infty}
\sum_{m'=-l'}^{l'}
g_{\rm B}^{(l'm')}\,
{}_{-2}Y_{l'm'}(\pi-\theta,\phi),
\end{align}
where 
\begin{align}
f_{\rm B}^{(l'm')}
&=
\frac{\pi e^{-i\Phi}}{i\omega}\sum_{l= 2}^{\infty}
\sum_{m=-l}^{l}
C_{l'}^{(l,m)}(|a|\omega)\,
D^{l'}_{m'm}(\gamma)\,
{}_{-2}S_{lm}(\gamma,0;|a|\omega)\Delta^{+}_{lm},
\\
g_{\rm B}^{(l'm')}
&=
\frac{\pi e^{-i\Phi}}{i\omega}\sum_{l=2}^{\infty}
\sum_{m=-l}^{l}
C_{l'}^{(l,m)}(|a|\omega)\,
(-1)^{m'+m}
D^{l'}_{m'm}(\gamma)\,
{}_{-2}S_{lm}(\gamma,0;|a|\omega)\, \Delta^{-}_{lm}\,.
\end{align}

We note that for a given spin order $n$, $C_{l'}^{(l,m)}(|a|\omega)$ is non-vanishing only when $|l-l'|\leq n$, and the infinite summation in these equations therefore reduces to a finite sum,
\begin{align}
f_{\rm B}^{(l'm')}\Big|_{|a|^n}
&=
\frac{\pi e^{-i\Phi}}{i\omega}\sum_{n_1+n_2+n_3=n}\sum_{l= {\rm Max}(2,|l'-n_{1}|)}^{l'+n_{1}}
\sum_{m=-l}^{l}\nn\\
&
\Big(C_{l'}^{(l,m)}(|a|\omega)\Big|_{|a|^{n_1}}\Big)\,
D^{l'}_{m'm}(\gamma)\,
\Big({}_{-2}S_{lm}(\gamma,0;|a|\omega)\Big|_{|a|^{n_2}}\Big)\Big(\Delta^{+}_{lm}\Big|_{|a|^{n_3}}\Big),
\\
g_{\rm B}^{(l'm')}\Big|_{|a|^n}
&=
\frac{\pi e^{-i\Phi}}{i\omega}\sum_{n_1+n_2+n_3=n}\sum_{l= {\rm Max}(2,|l'-n_{1}|)}^{l'+n_{1}}
\sum_{m=-l}^{l}\nn\\
&\Big(C_{l'}^{(l,m)}(|a|\omega)\Big|_{|a|^{n_1}}\Big)\,
(-1)^{m+m'}
D^{l'}_{m'm}(\gamma)\,
 \Big({}_{-2}S_{lm}(\gamma,0;|a|\omega)\Big|_{|a|^{n_2}}\Big)\Big(\Delta^{-}_{lm}\Big|_{|a|^{n_3}}\Big)\, .
\end{align}

\section{One-loop Calculation}\label{app:1loopRes}
In this appendix, we explicitly compute the one-loop contribution $\mathcal{M}_{\rmA,\text{fin.}}^{1\text{Loop}}\Big|_{\epsbar a^{6}}$. As discussed in the main text, this contribution is obtained by sewing the spinless Compton amplitude to the $\mathcal(\epsbar^0 |a|^6)$ contact term via graviton exchange, summing over the two possible orderings,
\begin{align}\label{eq:1loopeps1PieceAp}
\mathcal{M}_{\rmA, \text{fin.}}^{1\text{Loop}}\Big|_{\epsbar a^{6}}&\equiv\Big(\kern-0.6em\begin{tikzpicture}[scale=1.5, transform shape, baseline=0.1cm]
\tikzstyle{every node}=[font=\small]
\begin{feynman}
    \vertex (a) {};
    \vertex [right=0.6cm of a] (f2) [myblob2] {$\,\mathbf{0}\,$};
    \vertex [right=0.7cm of f2] (f3) [HV2] {$\,\mathbf{1}\,$};
    \vertex [above=0.3cm of f2]{\tiny $6$};
    \vertex [above=0.3cm of f3]{\tiny $0$};
    \vertex [right=1.3cm of f2] (c) {};
    \vertex [above=0.5cm of f2] (gm) {};
    \vertex [left=0.5cm of gm] (g2) {};
    \vertex [right=1.2cm of gm] (g20) {};
    \vertex [right=0.35cm of f2] (cut1);
    \vertex [below=0.15cm of cut1] (cut1u);
    \vertex [above=0.1cm of cut1] (cut1b);
    \vertex [above=0.4cm of cut1b] (cut1bb);
    \diagram* {
        (a) -- [fermion,thick] (f2) -- [fermion,thick] (f3)-- [fermion,thick] (c),
        (f3)-- [photon,ultra thick,momentum'=\(\ell\), out=145,in=35,looseness=0.5,min distance=0.2cm] (f2),
        (g2) -- [photon, ultra thick,postaction={
					decorate,
					decoration={
						markings,
						mark=at position 1.0 with
						{\arrow[scale=1., very thick]{Latex}}
					}
				}] (f2),
        (f3) -- [photon, ultra thick,postaction={
					decorate,
					decoration={
						markings,
						mark=at position 1.0 with
						{\arrow[scale=1., very thick]{Latex}}
					}
				}] (g20),
         (cut1u)--[red,thick] (cut1b),
      (cut1b)--[dashed,red,thick] (cut1bb)
    };
\end{feynman}
\end{tikzpicture}\kern-0.6em\Big)+\Big(\kern-0.6em\begin{tikzpicture}[scale=1.5, transform shape, baseline=0.1cm]
\tikzstyle{every node}=[font=\small]
\begin{feynman}
    \vertex (a) {};
    \vertex [right=0.6cm of a] (f2) [HV2] {$\,\mathbf{1}\,$};
    \vertex [right=0.7cm of f2] (f3) [myblob2] {$\,\mathbf{0}\,$};
    \vertex [above=0.3cm of f2]{\tiny $0$};
    \vertex [above=0.3cm of f3]{\tiny $6$};
    \vertex [right=1.3cm of f2] (c) {};
    \vertex [above=0.5cm of f2] (gm) {};
    \vertex [left=0.5cm of gm] (g2) {};
    \vertex [right=1.2cm of gm] (g20) {};
    \vertex [right=0.35cm of f2] (cut1);
    \vertex [below=0.15cm of cut1] (cut1u);
    \vertex [above=0.1cm of cut1] (cut1b);
    \vertex [above=0.4cm of cut1b] (cut1bb);
    \diagram* {
        (a) -- [fermion,thick] (f2) -- [fermion,thick] (f3)-- [fermion,thick] (c),
        (f3)-- [photon,ultra thick,momentum'=\(\ell\), out=145,in=35,looseness=0.5,min distance=0.2cm] (f2),
        (g2) -- [photon, ultra thick,postaction={
					decorate,
					decoration={
						markings,
						mark=at position 1.0 with
						{\arrow[scale=1., very thick]{Latex}}
					}
				}] (f2),
        (f3) -- [photon, ultra thick,postaction={
					decorate,
					decoration={
						markings,
						mark=at position 1.0 with
						{\arrow[scale=1., very thick]{Latex}}
					}
				}] (g20),
         (cut1u)--[red,thick] (cut1b),
      (cut1b)--[dashed,red,thick] (cut1bb)
    };
\end{feynman}
\end{tikzpicture}\kern-0.6em\Big)\, .
\end{align}
Here, the tree-level amplitudes we use are 
\begin{align}\label{eq:a6e0TreeConact}
  \CM_{\rmS,\text{fin.}}^{(1,0)}(k_1,k_2)
  &\equiv
  32\pi GM^2\frac{\left(v\cdot F_1\cdot F^*_2\cdot v\right)^2}
  {2\,p_1\cdot p_2\left(v\cdot p_1\right)^2}\Big|_{p_1\to k_1,p_2\to -k_2}^{\veps_2^*\to \veps_2},\nn\\
 \mathcal{M}^{(0,6)}_{\rmA,\rm c}(k_1,k_2)&\equiv \mathcal{M}^{(0,6)}_{\rmA,\rm c}\Big|_{p_1\to k_1,p_2\to -k_2}^{\veps_2^*\to \veps_2}\, ,
\end{align}
with $k_1, k_2$ taken as incoming momenta. By gluing the two tree amplitudes and imposing the velocity-cut condition $\delta\left(2M v\cdot(\ell+p_1)\right)$, the one-loop contribution in dimensional regularization and without dropping the infrared factor is
\begin{equation}\label{eq:integrand}
    \begin{aligned}
        &\mathcal{M}_{\rmA}^{1\text{Loop}}\Big|_{\epsbar a^{6}}
        =
        \tilde\mu^{2\epsilon}\int \frac{d^{D}\ell}{(2\pi)^D}
        \frac{2\pi\delta\left(2Mv\cdot(\ell+p_1)\right)}{\ell^2}
        \\
        &\qquad\times
        \bigg[
        \sum_{\rm pols.}
    \CM_{\rmS,\text{fin.}}^{(1,0)}(p_1,\ell)
        \CM_{\rmA,c}^{(0,6)}(-\ell,-p_2)
        +
        \sum_{\rm pols.}
        \CM_{\rmA,c}^{(0,6)}(p_1,\ell)
    \CM_{\rmS,\text{fin.}}^{(1,0)}(-\ell,-p_2)
        \bigg]\, ,
    \end{aligned}
\end{equation}
where $\tilde\mu^2= {e^{\gamma_E}\mu^2\over 4\pi}$ and $D=4-2\epsilon$. Summing over the intermediate graviton states \cite{Kosmopoulos:2020pcd} and repeatedly using the gauge-invariant identity,
\begin{equation}
    (F\mdot Y)^\mu = \frac{(p_1\mdot Y)(F_i\mdot v)^\mu - (Y\mdot F_i\mdot v)p_i^\mu}{v\mdot p_i}, \quad (F_i\mdot v)^\mu\equiv F_i^{\mu\nu}v_{\nu},
\end{equation}
we can rewrite the following dot products $\ell \cdot F_1 \cdot a,\, \ell \cdot F_2 \cdot a,\, \ell \cdot F_1 \cdot F_2 \cdot v$, and $\ell \cdot F_2 \cdot F_1 \cdot v$ in terms of scalar products among our kinematic vectors $X_i\cdot X_j$, $X_i\in\{a^\mu,v^\mu, p_1^\mu, p_2^\mu, \ell^\mu\}$, and $\ell\mdot F_i \mdot v$, $i=1,2$. The full integrand can therefore be expressed in terms of the integral family,
\begin{equation}
   K_{\nu_1,\,\nu_2,\,\nu_3,\,\lambda_1,\,\lambda_2,\,\lambda_3}
   =
   \tilde\mu^{2\epsilon}
   \int \frac{d^{D}\ell}{(2\pi)^D}
   \frac{
   2\pi\delta\left(v\cdot(\ell+p_1)\right)
   (a\mdot\ell)^{\lambda_1}
   (\ell\mdot F_1\mdot v)^{\lambda_2}
   (\ell\mdot F_2\mdot v)^{\lambda_3}
   }
   {(\ell^2)^{\nu_1}
   (\ell\mdot p_1)^{\nu_2}
   (\ell\mdot p_2)^{\nu_3}}.
\end{equation}
We reduce this family using integration-by-parts identities implemented in the LiteRed2 package \cite{Lee:2012cn,Lee:2013mka}. The numerator structures $(a\mdot\ell)$, $(\ell\mdot F_1\mdot v)$, and $(\ell\mdot F_2\mdot v)$ are included directly in the integration-by-parts basis. We find that two master integrals span this family.
\begin{equation}
    \mathcal{K}_{\rm bub.}\equiv K_{1,0,0,0,0,0}, \quad \mathcal{K}_{\rm box.}\equiv K_{1,1,1,0,0,0}.
\end{equation}
However, after assembling the full amplitude, the cut-box master integral coefficient vanishes; thus, only the cut-bubble integral contributes. Its value is
\begin{equation}
\mathcal{K}_{\rm bub.}
=
\tilde\mu^{2\epsilon}
\int \frac{d^{D}\ell}{(2\pi)^D}
\frac{2\pi\delta\left((\ell+p_1)\mdot v\right)}{\ell^2}
=
-\frac{i\omega}{4\pi}
\Bigg[
1+\epsilon\bigg(i\pi+2-2\gamma_E\bigg)
\Bigg]
+\mathcal{O}(\epsilon^2),
\end{equation}
where we have fixed the renormalization scale to $\mu=2\euler^{-\gamma_{\rm E}}\omega$. Substituting the master integral back into the amplitude yields a one-loop amplitude with an infrared divergence that is proportional to the lower-order amplitude, as implied by Weinberg's theorem \cite{Weinberg:1965nx}. This provides a useful consistency check. Removing the divergent phase factor yields the finite amplitude,
\begin{equation}
\begin{aligned}
\mathcal{M}_{\rmA,\text{fin.}}^{1\text{Loop}}\Big|_{\epsbar a^{6}}&= \Big[\mathcal{M}_{\rmA}^{1\text{Loop}}\Big|_{\epsbar a^{6}}
    +\frac{i \epsbar}{\epsilon}\CM_{\rmA,\rm c}^{(0,6)}\Big]\Big|_{\epsilon\to 0}\, \\
    &=\epsbar\,\big(\pi +i (2 \gamma_{\rm E} -{25\over 12})\big) \mathcal{M}^{(0,6)}_{\rmA,\rm c}\,
    \\
    &+32\pi GM^2 i \,a\mdot p_1\, a\mdot p_2 \,\mathsf{c}_1 \left(2 U_{10}-U_{22}-2 U_{27}\right)+32\pi GM^2 i \,\omega^2 \,|a| ^2\times\\
    &
   \Bigg(-\frac{1}{6}\Big[
\left(3\mathsf{c}_1+14\mathsf{c}_2+3\mathsf{c}_3\right)U_{21}
+\left(\mathsf{c}_1+2\mathsf{c}_2+\mathsf{c}_3\right)U_{22}+\mathsf{c}_2
 \left(4U_{27}-U_{10}\right)
 \Big]
\\
&\quad
+\frac{1}{12}\Big[
\left(14\mathsf{c}_2+3\mathsf{c}_3\right)U_{11}
-\mathsf{c}_1
\left(11U_{9}-12U_{25}+16U_{26}\right)
 \Big]
\\
&\quad
+\frac{1}{24}
 \left(7\mathsf{c}_1-8\mathsf{c}_2-7\mathsf{c}_3\right)
 \left(U_{13}+U_{14}\right)-\frac{1}{48}
 \left(\mathsf{c}_1+\mathsf{c}_3\right)
\left(U_{12}-16 U_{27}+12U_{28}\right)
\Bigg),
\end{aligned}
\end{equation}
where 
\begin{align}
U_{25} &= \left(a\mdot F_1\mdot v\right)^2
\left(a\mdot F_2^*\mdot v\right)^2,&
U_{26} &= a\mdot F_1\mdot F_2^*\mdot a\,
a\mdot F_1\mdot v\,
a\mdot F_2^*\mdot v,\nn\\
U_{27} &= |a|^2\,
a\mdot F_1\mdot F_2^*\mdot a\,
v\mdot F_1\mdot F_2^*\mdot v,&
U_{28} &= |a|^2\,
\operatorname{tr}(F_1\mdot F_2^*)\,
a\mdot F_1\mdot F_2^*\mdot a\, .
\end{align} 
We have confirmed that the four $U$-functions contain only $l=2$ modes. 
\section{Data for Contact Term Determination}\label{app:fgdata}
In this appendix, we summarize the black hole perturbation theory and effective field theory amplitude information used in Section~\ref{sec:contactTerms} to determine the contact terms. For the helicity-conserving scattering function from black hole perturbation theory and the helicity-conserving component of $\mathcal M_{\rmS,\rm fin.}$ at order $\terms{\epsbar|a|^{5}}$, we need the $l=2$ modes,
\begin{align}
f_{\rm B}^{(2,-2)}\Big|_{\epsbar|a|^{5}}
&=
\frac{109}{1512}\,\sqrt{5\pi}\,\epsbar\,|a|^{5}\,\omega^{4}\,
\sin^{4}(\gamma)\cos(\gamma),
\nn\\[6pt]
f_{\rm B}^{(2,-1)}\Big|_{\epsbar|a|^{5}}
&=
-\frac{1}{60480}\sqrt{\frac{\pi}{5}}\,\epsbar\,|a|^{5}\,\omega^{4}\,\sin^{3}(\gamma)\nn\\
&\qquad\times\Big(
17775\cos(2\gamma)
-645120\zeta(5)
+677376\zeta(3)
-68623
\Big),
\nn\\[6pt]
f_{\rm B}^{(2,0)}\Big|_{\epsbar|a|^{5}}
&=
\frac{1}{2520}\sqrt{\frac{\pi}{30}}\,\epsbar\,|a|^{5}\,\omega^{4}\,
\sin^{2}(\gamma)\cos(\gamma)\nn\\
&\qquad\times\Big(
1875\cos(2\gamma)
-161280\zeta(5)
+169344\zeta(3)
-20627
\Big),
\nn\\[6pt]
f_{\rm B}^{(2,1)}\Big|_{\epsbar|a|^{5}}
&=
-\frac{1}{30240}\sqrt{\frac{\pi}{5}}\,\epsbar\,|a|^{5}\,\omega^{4}\,\sin(\gamma)\nn\\
&\qquad\times\Big(
1605\cos(4\gamma)
+16\big(31752\zeta(3)-30240\zeta(5)-3901\big)\cos(2\gamma)
\nn\\
&\hspace{4.0cm}
-849408\zeta(5)
+857472\zeta(3)
-96213
\Big),
\nn\\[6pt]
f_{\rm B}^{(2,2)}\Big|_{\epsbar|a|^{5}}
&=
\frac{1}{60480}\sqrt{\frac{\pi}{5}}\,\epsbar\,|a|^{5}\,\omega^{4}\,\cos(\gamma)\nn\\
&\qquad\times\Big(
1395\cos(4\gamma)
+28\big(24192\zeta(3)-23040\zeta(5)-3121\big)\cos(2\gamma)
\nn\\
&\hspace{4.0cm}
-2107392\zeta(5)
+2075136\zeta(3)
-215191
\Big),
\end{align}

\begin{align}
\mathbb{P}^{(2,-2)}_f\Big[\mathcal{M}_{\rmS, \text{fin.}}\Big]\Big|_{\epsbar|a|^5}
&=
\frac{109}{1512}\sqrt{5\pi}\,\epsbar\,|a|^{5}\,\omega^4
\cos(\gamma)\sin^4(\gamma),
\nn\\[6pt]
\mathbb{P}^{(2,-1)}_f\Big[\mathcal{M}_{\rmS, \text{fin.}}\Big]\Big|_{\epsbar|a|^5}
&=
-\frac{1}{60480}\sqrt{\frac{\pi}{5}}\,\epsbar\,|a|^{5}\,\omega^4
\left(
    20081 + 17775\cos(2\gamma)
\right)\sin^3(\gamma),
\nn\\[6pt]
\mathbb{P}^{(2,0)}_f\Big[\mathcal{M}_{\rmS, \text{fin.}}\Big]\Big|_{\epsbar|a|^5}
&=
\frac{1}{5040}\sqrt{\frac{\pi}{30}}\,\epsbar\,|a|^{5}\,\omega^4
\left(
    4973\cos(\gamma) + 1875\cos(3\gamma)
\right)\sin^2(\gamma),\nn
\\[6pt]
\mathbb{P}^{(2,1)}_f\Big[\mathcal{M}_{\rmS, \text{fin.}}\Big]\Big|_{\epsbar|a|^5}
&=
\frac{1}{60480}\sqrt{\frac{\pi}{5}}\,\epsbar\,|a|^{5}\,\omega^4
\left(
    1658\sin(\gamma)
    - 2507\sin(3\gamma)
    - 1605\sin(5\gamma)
\right),
\nn\\[6pt]
\mathbb{P}^{(2,2)}_f\Big[\mathcal{M}_{\rmS, \text{fin.}}\Big]\Big|_{\epsbar|a|^5}
&=
\frac{1}{120960}\sqrt{\frac{\pi}{5}}\,\epsbar\,|a|^{5}\,\omega^4
\left(
    -4362\cos(\gamma)
    + 2711\cos(3\gamma)
    + 1395\cos(5\gamma)
\right).
\end{align}
At order $\terms{\epsbar|a|^{5}}$, the helicity-conserving components of $\mathcal M_{\rmS,\rm fin.}$ for $l\geq 3$ and the helicity-reversing components for $l\geq 2$ coincide with the corresponding modes of the scattering functions in black hole perturbation theory. We have checked this explicitly up to $l=6$ and provide some examples,
\begin{align}
&\mathbb{P}^{(2,-2)}_g\Big[\mathcal{M}_{\rmS, \text{fin.}}\Big]\Big|_{\epsbar|a|^5}=g_{\rm B}^{(2,-2)}\Big|_{\epsbar|a|^{5}}
=
\frac{\sqrt{\frac{\pi}{5}}\,\epsbar\,|a|^{5}\,\omega^4}{1512}
\cos(\gamma)\sin^4(\gamma),
\nn\\[6pt]
&\mathbb{P}^{(2,-1)}_g\Big[\mathcal{M}_{\rmS, \text{fin.}}\Big]\Big|_{\epsbar|a|^5}=g_{\rm B}^{(2,-1)}\Big|_{\epsbar|a|^{5}}
=
\frac{\sqrt{\frac{\pi}{5}}\,\epsbar\,|a|^{5}\,\omega^4}{3024}
\left(
    11 + 9\cos(2\gamma)
\right)\sin^3(\gamma),
\nn\\[6pt]
&\mathbb{P}^{(2,0)}_g\Big[\mathcal{M}_{\rmS, \text{fin.}}\Big]\Big|_{\epsbar|a|^5}=g_{\rm B}^{(2,0)}\Big|_{\epsbar|a|^{5}}
=
\frac{\sqrt{\frac{5\pi}{6}}\,\epsbar\,|a|^{5}\,\omega^4}{504}
\cos(\gamma)
\left(
    5 + 3\cos(2\gamma)
\right)\sin^2(\gamma),\nn
\\[6pt]
&\mathbb{P}^{(2,1)}_g\Big[\mathcal{M}_{\rmS, \text{fin.}}\Big]\Big|_{\epsbar|a|^5}=g_{\rm B}^{(2,1)}\Big|_{\epsbar|a|^{5}}
=
\frac{\sqrt{5\pi}\,\epsbar\,|a|^{5}\,\omega^4}{48384}
\left(
    170\sin(\gamma)
    + 187\sin(3\gamma)
    + 33\sin(5\gamma)
\right),
\nn\\[6pt]
&\mathbb{P}^{(2,2)}_g\Big[\mathcal{M}_{\rmS, \text{fin.}}\Big]\Big|_{\epsbar|a|^5}=g_{\rm B}^{(2,2)}\Big|_{\epsbar|a|^{5}}
=
\frac{\sqrt{\frac{\pi}{5}}\,\epsbar\,|a|^{5}\,\omega^4}{12096}
\cos(\gamma)
\left(
    857
    + 836\cos(2\gamma)
    + 99\cos(4\gamma)
\right),
\label{eq:gB-l2}
\end{align}

\begin{equation}
    \begin{aligned}
 \mathbb{P}^{(3,-3)}_f\Big[\mathcal{M}_{\rmS, \text{fin.}}\Big]\Big|_{\epsbar|a|^5}&=f_{\rm B}^{(3,-3)}\Big|_{\epsbar|a|^5}=-{1217\over 3600} \sqrt{\frac{\pi }{42}} \epsbar\,|a|^{5}\,  \omega ^4 \sin ^5(\gamma ),
   \\[4pt]
  \mathbb{P}^{(3,-2)}_f\Big[\mathcal{M}_{\rmS, \text{fin.}}\Big]\Big|_{\epsbar|a|^5}&=f_{\rm B}^{(3,-2)}\Big|_{\epsbar|a|^5}=\frac{41}{135} \sqrt{\frac{\pi }{7}} \epsbar\,|a|^{5}\,  \omega ^4 \sin ^4(\gamma ) \cos
   (\gamma ), \\[4pt]
  \mathbb{P}^{(3,-1)}_f\Big[\mathcal{M}_{\rmS, \text{fin.}}\Big]\Big|_{\epsbar|a|^5}&=f_{\rm B}^{(3,-1)}\Big|_{\epsbar|a|^5}=-\frac{1}{432} \sqrt{\frac{\pi }{70}} \epsbar\,|a|^{5}\,  \omega ^4 \sin ^3(\gamma )
   \big(285 \cos (2 \gamma )+179\big) ,\\[4pt]
  \mathbb{P}^{(3,0)}_f\Big[\mathcal{M}_{\rmS, \text{fin.}}\Big]\Big|_{\epsbar|a|^5}&=f_{\rm B}^{(3,0)}\Big|_{\epsbar|a|^5}=\frac{1}{144} \sqrt{\frac{\pi }{210}} \epsbar\,|a|^{5}\,  \omega ^4 \sin ^2(\gamma ) \big(77
   \cos (\gamma )+83 \cos (3 \gamma )\big), \\[4pt]
   \mathbb{P}^{(3,1)}_f\Big[\mathcal{M}_{\rmS, \text{fin.}}\Big]\Big|_{\epsbar|a|^5}&=f_{\rm B}^{(3,1)}\Big|_{\epsbar|a|^5}=-{\sqrt{\frac{\pi }{70}} \epsbar |a|^{5}\,  \omega ^4\over 6912} \big(170 \sin (\gamma )-365 \sin
   (3 \gamma )+441 \sin (5 \gamma )\big), \\[4pt]
   \mathbb{P}^{(3,2)}_f\Big[\mathcal{M}_{\rmS, \text{fin.}}\Big]\Big|_{\epsbar|a|^5}&=f_{\rm B}^{(3,2)}\Big|_{\epsbar|a|^5}={\sqrt{\frac{\pi }{7}} \epsbar\,|a|^{5}\,  \omega ^4 \over 43200}\big(210 \cos (\gamma )-455 \cos (3
   \gamma )+501 \cos (5 \gamma )\big), \\[4pt]
   \mathbb{P}^{(3,3)}_f\Big[\mathcal{M}_{\rmS, \text{fin.}}\Big]\Big|_{\epsbar|a|^5}&=f_{\rm B}^{(3,3)}\Big|_{\epsbar|a|^5}={\sqrt{\frac{\pi }{42}} \epsbar\,|a|^{5}\,  \omega ^4\over 5760} \sin (\gamma ) \big(-44 \cos (2
   \gamma )+127 \cos (4 \gamma )+45\big). \\
    \end{aligned}
\end{equation}

At order $\terms{\epsbar|a|^{6}}$, we show the difference between black hole perturbation theory and the amplitude $\mathcal M_{\rmS, \rm fin.}$ for both helicity-conserving and helicity-reversing modes at $l=2$,
\begin{equation}
\begin{aligned}
f_{\rm diff.}^{(2,-2)}\Big|_{\epsbar |a|^6}
&=
\frac{|a|^6\epsbar\,\omega^5}{4725}
\sqrt{\frac{\pi}{5}}\,
\sin^4(\gamma)
\bigl(
5717 + 19830\zeta(2)
\nn\\
&\quad
-182436\zeta(4)
+158640\zeta(6)
-1260(2\gamma_{\rm E}-i\pi)
\bigl(
-1+10\zeta(2)-92\zeta(4)+80\zeta(6)
\bigr)
\nn\\
&\quad
-129360\zeta(3)
+837200\zeta(5)
-705600\zeta(7)
\bigr),
\nn\\[8pt]
f_{\rm diff.}^{(2,-1)}\Big|_{\epsbar |a|^6}
&=
\frac{2|a|^6\epsbar\omega^5}{4725}
\sqrt{\frac{\pi}{5}}\,
\cos(\gamma)\sin^3(\gamma)
\bigl(
-3407-19830\zeta(2)
\nn\\
&\quad
+182436\zeta(4)
-158640\zeta(6)
+1260(2\gamma_{\rm E}-i\pi)
\bigl(
-1+10\zeta(2)-92\zeta(4)+80\zeta(6)
\bigr),
\nn\\
&\quad
+111720\zeta(3)
-820400\zeta(5)
+705600\zeta(7)
\bigr),
\nn\\[8pt]
f_{\rm diff.}^{(2,0)}\Big|_{\epsbar |a|^6}
&=
\frac{|a|^6\epsbar\omega^5}{9450}
\sqrt{\frac{\pi}{30}}\,
\sin^2(\gamma)
\bigl(
65359
+8820(2\gamma_{\rm E}-i\pi)
\nn\\
&\quad
+356940\zeta(2)
-3379032\zeta(4)
+2982432\zeta(6)
\nn\\
&\quad
-5040(2\gamma_{\rm E}-i\pi)
\bigl(
45\zeta(2)-426\zeta(4)+376\zeta(6)
\bigr)
\nn\\
&\quad
-2248960\zeta(3)
+15537760\zeta(5)
-13265280\zeta(7)
\nn\\
&\quad
-6\cos(2\gamma)
\bigl(
-1097-19830\zeta(2)
+182436\zeta(4)
-158640\zeta(6)
\nn\\
&\qquad
+1260(2\gamma_{\rm E}-i\pi)
\bigl(
-1+10\zeta(2)-92\zeta(4)+80\zeta(6)
\bigr)
\nn\\
&\qquad
+94080\zeta(3)
-803600\zeta(5)
+705600\zeta(7)
\bigr)
\bigr),
\nn\\[8pt]
f_{\rm diff.}^{(2,1)}\Big|_{\epsbar |a|^6}
&=
\frac{|a|^6\epsbar\omega^5}{18900}
\sqrt{\frac{\pi}{5}}\,
\bigl(
\sin(4\gamma)
\bigl(
1213-19830\zeta(2)
+182436\zeta(4)
\nn\\
&\qquad
-158640\zeta(6)
+1260(2\gamma_{\rm E}-i\pi)
\bigl(
-1+10\zeta(2)-92\zeta(4)+80\zeta(6)
\bigr)
\nn\\
&\qquad
+76440\zeta(3)
-786800\zeta(5)
+705600\zeta(7)
\bigr)
\nn\\
&\quad
+\sin(2\gamma)
\bigl(
-49911-277620\zeta(2)
+2649288\zeta(4)
-2347872\zeta(6)
\nn\\
&\qquad
+1260(2\gamma_{\rm E}-i\pi)
\bigl(
-3+140\zeta(2)-1336\zeta(4)+1184\zeta(6)
\bigr)
\nn\\
&\qquad
+1764560\zeta(3)
-12213600\zeta(5)
+10442880\zeta(7)
\bigr)
\bigr),
\nn\\[8pt]
f_{\rm diff.}^{(2,2)}\Big|_{\epsbar |a|^6}
&=
\frac{|a|^6\epsbar\omega^5}{113400}
\sqrt{\frac{\pi}{5}}\,
\bigl(
-12600(2\gamma_{\rm E}-i\pi)
\nn\\
&\quad\times
\bigl(
105\zeta(2)-1014\zeta(4)+904\zeta(6)
\bigr)
\nn\\
&\quad
+5\bigl(
71267
+8820(2\gamma_{\rm E}-i\pi)
+416430\zeta(2)
-4021524\zeta(4)
\nn\\
&\qquad
+3585264\zeta(6)
-2679152\zeta(3)
+18616976\zeta(5)
-15946560\zeta(7)
\bigr)
\nn\\
&\quad
-3\cos(4\gamma)
\bigl(
3523-19830\zeta(2)
+182436\zeta(4)
-158640\zeta(6)
\nn\\
&\qquad
+1260(2\gamma_{\rm E}-i\pi)
\bigl(
-1+10\zeta(2)-92\zeta(4)+80\zeta(6)
\bigr)
\nn\\
&\qquad
+58800\zeta(3)
-770000\zeta(5)
+705600\zeta(7)
\bigr)
\nn\\
&\quad
-6\cos(2\gamma)
\bigl(
-38851-277620\zeta(2)
+2649288\zeta(4)
-2347872\zeta(6)
\\
&\qquad
+1260(2\gamma_{\rm E}-i\pi)
\bigl(
-3+140\zeta(2)-1336\zeta(4)+1184\zeta(6)
\bigr)
\\
&\qquad
+1656480\zeta(3)
-12103840\zeta(5)
+10442880\zeta(7)
\bigr)
\bigr),
\end{aligned}
\end{equation}

\begin{align}
g_{\mathrm{diff.}}^{(2,-2)}
\Big|_{\bar{\epsilon}|a|^{6}}
&=
\frac{1}{15}\sqrt{\frac{\pi}{5}}\,
\bar{\epsilon}\,\omega^{5}|a|^{6}\sin^{4}(\gamma)
\left(
1-10\zeta(2)+92\zeta(4)-80\zeta(6)
\right),
\nn\\[1ex]
g_{\mathrm{diff.}}^{(2,-1)}
\Big|_{\bar{\epsilon}|a|^{6}}
&=
-\frac{2}{15}\sqrt{\frac{\pi}{5}}\,
\bar{\epsilon}\,\omega^{5}|a|^{6}
\cos(\gamma)\sin^{3}(\gamma)
\left(
-1+10\zeta(2)-92\zeta(4)+80\zeta(6)
\right),
\nn\\[1ex]
g_{\mathrm{diff.}}^{(2,0)}
\Big|_{\bar{\epsilon}|a|^{6}}
&=
\frac{1}{30}\sqrt{\frac{\pi}{30}}\,
\bar{\epsilon}\,\omega^{5}|a|^{6}\sin^{2}(\gamma)
\Big(
7-180\zeta(2)+1704\zeta(4)-1504\zeta(6)
\nonumber\\
&\hspace{4.4cm}
+6\cos(2\gamma)
\left(
1-10\zeta(2)+92\zeta(4)-80\zeta(6)
\right)
\Big),
\nn\\[1ex]
g_{\mathrm{diff.}}^{(2,1)}
\Big|_{\bar{\epsilon}|a|^{6}}
&=
\frac{1}{60}\sqrt{\frac{\pi}{5}}\,
\bar{\epsilon}\,\omega^{5}|a|^{6}\sin(2\gamma)
\Big(
3-140\zeta(2)+1336\zeta(4)-1184\zeta(6)
\nonumber\\
&\hspace{4.4cm}
+2\cos(2\gamma)
\left(
1-10\zeta(2)+92\zeta(4)-80\zeta(6)
\right)
\Big),
\nn\\[1ex]
g_{\mathrm{diff.}}^{(2,2)}
\Big|_{\bar{\epsilon}|a|^{6}}
&=
\frac{1}{360}\sqrt{\frac{\pi}{5}}\,
\bar{\epsilon}\,\omega^{5}|a|^{6}
\Big(
\cos(2\gamma)
\left(
18-840\zeta(2)+8016\zeta(4)-7104\zeta(6)
\right)
\nonumber\\
&\hspace{2.8cm}
+5\left(
7-210\zeta(2)+2028\zeta(4)-1808\zeta(6)
\right)
\nonumber\\
&\hspace{2.8cm}
+3\cos(4\gamma)
\left(
1-10\zeta(2)+92\zeta(4)-80\zeta(6)
\right)
\Big).
\end{align}
 At $l=3$, we show the difference between black hole perturbation theory and $\mathcal M_{\rmS,\rm fin.}$ for the helicity-conserving modes, 
\begin{align}
f_{\mathrm{diff.}}^{(3,-3)}
\Big|_{\bar{\epsilon}|a|^{6}}
&= 0, \qquad
f_{\mathrm{diff.}}^{(3,-2)}
\Big|_{\bar{\epsilon}|a|^{6}}
=
\frac{4}{135}\sqrt{\frac{\pi}{7}}\,
\bar{\epsilon}\,\omega^{5}|a|^{6}\sin^{4}(\gamma)
\left(
-11+84\zeta(3)-80\zeta(5)
\right),
\nn\\[1ex]
f_{\mathrm{diff.}}^{(3,-1)}
\Big|_{\bar{\epsilon}|a|^{6}}
&=
-\frac{4}{27}\sqrt{\frac{2\pi}{35}}\,
\bar{\epsilon}\,\omega^{5}|a|^{6}
\cos(\gamma)\sin^{3}(\gamma)
\left(
-11+84\zeta(3)-80\zeta(5)
\right),
\nn\\[1ex]
f_{\mathrm{diff.}}^{(3,0)}
\Big|_{\bar{\epsilon}|a|^{6}}
&=
\frac{2}{135}\sqrt{\frac{2\pi}{105}}\,
\bar{\epsilon}\,\omega^{5}|a|^{6}\sin^{2}(\gamma)
\Big(
-155+1268\zeta(3)-1232\zeta(5)
\nonumber\\
&\hspace{5.6cm}
+15\cos(2\gamma)
\left(
-11+84\zeta(3)-80\zeta(5)
\right)
\Big),
\nn\\[1ex]
f_{\mathrm{diff.}}^{(3,1)},
\Big|_{\bar{\epsilon}|a|^{6}}
&=
-\frac{2}{135}\sqrt{\frac{2\pi}{35}}\,
\bar{\epsilon}\,\omega^{5}|a|^{6}\sin(2\gamma)
\Big(
428\zeta(3)-9\left(5+48\zeta(5)\right)
\nonumber\\
&\hspace{5.4cm}
+5\cos(2\gamma)
\left(
-11+84\zeta(3)-80\zeta(5)
\right)
\Big),
\nn\\[1ex]
f_{\mathrm{diff.}}^{(3,2)}
\Big|_{\bar{\epsilon}|a|^{6}}
&=
\frac{1}{810}\sqrt{\frac{\pi}{7}}\,
\bar{\epsilon}\,\omega^{5}|a|^{6}
\Big(
-191+1796\zeta(3)-1808\zeta(5)
\nonumber\\
&\hspace{3.7cm}
+15\cos(4\gamma)
\left(
-11+84\zeta(3)-80\zeta(5)
\right)
\nonumber\\
&\hspace{3.7cm}
+12\cos(2\gamma)
\left(
428\zeta(3)-9\left(5+48\zeta(5)\right)
\right)
\Big),
\nn\\[1ex]
f_{\mathrm{diff.}}^{(3,3)}
\Big|_{\bar{\epsilon}|a|^{6}}
&=
\frac{1}{135}\sqrt{\frac{2\pi}{21}}\,
\bar{\epsilon}\,\omega^{5}|a|^{6}\sin(2\gamma)
\Big(
-79+772\zeta(3)-784\zeta(5)
\nonumber\\
&\hspace{5.4cm}
+3\cos(2\gamma)
\left(
-11+84\zeta(3)-80\zeta(5)
\right)
\Big).
\end{align}
Except for the modes shown above, $\mathcal M_{\rmS,\rm fin.}$ agrees with the scattering functions from black hole perturbation theory to order $\terms{\epsbar|a|^{6}}$. 

\bibliographystyle{JHEP}
\bibliography{KinematicAlgebra2}
\end{document}